\documentclass[sn-aps]{sn-jnl}

\usepackage{hyperref}
\hypersetup{linkcolor=blue,citecolor=red}

\newcommand{\hnn}[1]{\widehat{\boldsymbol #1}}
\newcommand{\medio}[1]{\langle #1\rangle}

\newcommand\beq{\begin{equation}}
\newcommand\eeq{\end{equation}}
\newcommand\beqa{\begin{eqnarray}}
\newcommand\eeqa{\end{eqnarray}}
\def\bal#1\eal{\begin{align}#1\end{align}}
\def\be{\begin{equation}}
\def\ee#1{\label{#1}\end{equation}}
\newcommand{\nn}{\nonumber\\}
\newcommand{\bnabla}{\boldsymbol{\nabla}}
\newcommand{\nuM}{\nu_{\text{M}}}
\newcommand{\hs}{\text{HS}}

\def\bc{\mathbf{v}}
\def\bcc{\mathbf{c}}
\def\bww{\mathbf{w}}
\def\bV{\mathbf{V}}
\def\ka{\kappa}
\def\bt{\widetilde\beta}
\def\at{\widetilde\alpha}
\def\bg{\mathbf{g}}
\def\bk{\hnn{\sigma}}
\def\a{{\alpha }}
\def\b{{\beta }}
\def\d{\sigma }
\def\bu{\overline{\mathbf{g}}}
\def\bx{\mathbf{r}}

\def\bw{\boldsymbol{\omega}}
\def\bW{\boldsymbol{\Omega}}
\def\bxi{\boldsymbol{\xi}}
\def\bJ{\mathbf{Q}}

\jyear{2022}%

\theoremstyle{thmstyleone}%
\theoremstyle{thmstyletwo}%

\theoremstyle{thmstylethree}%

\begin{document}

\title[Inelastic and rough Maxwell particles]{Granular gas  of inelastic and rough Maxwell particles}


\author[1]{\fnm{Gilberto M.} \sur{Kremer}}\email{kremer@fisica.ufpr.br}
\equalcont{These authors contributed equally to this work.}
\author*[2]{\fnm{Andr\'es} \sur{Santos}}\email{andres@unex.es}
\equalcont{These authors contributed equally to this work.}

\affil[1]{\orgdiv{Departamento de F\'{\i}sica}, \orgname{Universidade Federal do Paran\'a},
\orgaddress{
\city{Curitiba},
\country{Brazil}}}

\affil[2]{\orgdiv{Departamento de F\'{\i}sica and Instituto de Computaci\'on Cient\'ifica Avanzada (ICCAEx)}, \orgname{Universidad de Extremadura},
\orgaddress{
\city{Badajoz}, \postcode{E-06006},
\country{Spain}}}



\abstract{The most widely used model for granular gases is perhaps the inelastic hard-sphere model (IHSM), where the grains are assumed to be perfectly smooth spheres colliding with a constant coefficient of normal restitution. A much more tractable model is the inelastic Maxwell model (IMM), in which the velocity-dependent collision rate is replaced by an effective mean-field constant. This simplification has been taken advantage of by many researchers to find a number of exact results within the IMM. On the other hand, both the IHSM and IMM neglect the impact of  roughness---generally present in real grains---on the dynamic properties of a granular gas. This is remedied by the inelastic rough hard-sphere model (IRHSM), where, apart from the coefficient of normal restitution, a constant coefficient of tangential restitution is introduced. In parallel to the simplification carried out when going from the IHSM to the IMM, we propose in this paper an inelastic rough Maxwell model (IRMM) as a simplification of the IRHSM. The tractability of the proposed model is illustrated by the exact evaluation of the collisional moments  of first and second degree, and the most relevant ones of  third and fourth degree. The results are applied to the evaluation of the rotational-to-translational temperature ratio and the velocity cumulants in the  homogeneous cooling state. }




\maketitle

\section{Introduction}\label{sec1}

Granular gases are typically modeled as a system of agitated inelastic hard spheres \cite{BP04,G19,K10a}. In the basic inelastic hard-sphere model (IHSM) of granular gases, the particles are assumed to be smooth (i.e., with no rotational degrees of freedom) and the collisions are characterized by a constant coefficient of \emph{normal} restitution, $0<\alpha\leq 1$. If the number density of the gas is low enough, the most relevant physical quantity is the one-particle velocity distribution function (VDF), which obeys the  Boltzmann equation appropriately adapted to incorporate the inelastic nature of collisions. Nevertheless, the fact that the collision rate in the IHSM is proportional to the relative velocity of the two colliding particles prevents the associated collisional moments from being expressible in terms of a finite number of velocity moments, thus hampering the possibility of deriving analytical results.

The above difficulty is also present for molecular gases (where collisions are elastic, i.e., $\a=1$). In that case, it can be overcome by assuming that the gas is made of Maxwell molecules, that is, particles interacting via
a repulsive force inversely proportional to the fifth power of distance
\cite{M67,TM80,E81,GS03,S09b,K10a}. This makes the collision rate independent of the relative velocity, so that the collisional moments become bilinear combinations of velocity moments of the same or lower degree.
If collisions are inelastic ($\a<1$), one can still construct the so-called inelastic Maxwell model (IMM) by assuming an effective mean-field collision rate independent of the velocity
\cite{BCG00,CCG00,BK00,C01b,BK03,BC03,BCT03,SE03,G03a,BE04,GA05,BCT06,BG06,BC07,CT07,G07,BCG08,BCG09,CCC09,GT10,FPTT10,BGM10,KGS14,%
GG19,KG20,BMP02,BK02,KB02,EB02a,EB02b,S03,SG07,EB02c,GS07,GS11,SG12}.

\begin{figure}
\centering
\includegraphics[width=0.7\textwidth]{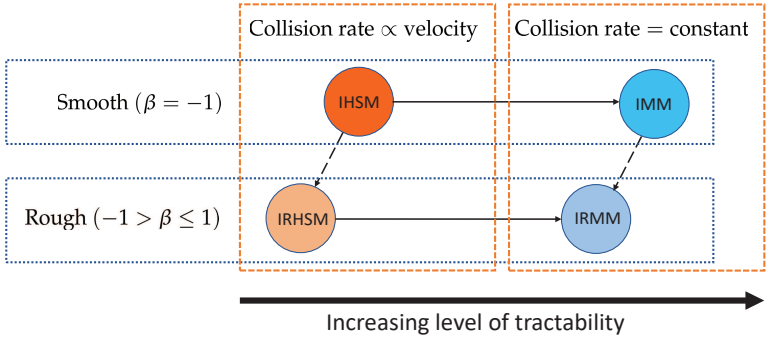}
\caption{Sketch of four granular-gas models: IHSM, IMM, IRHSM, and IRMM.}\label{sketch}
\end{figure}

The original IHSM and the simpler IMM capture many of the most relevant features of granular gases \cite{G03,G19}. On the other hand, both models leave out the roughness of real grains, which gives rise to frictional collisions and energy transfer between the translational and rotational degrees of freedom. A convenient way of modeling this roughness effect is by augmenting the IHSM by means of a constant coefficient of \emph{tangential} restitution $-1\leq\beta\leq 1$
\cite{JR85a,LS87,C89,L91,LB94,GS95,L95,L96,ZTPSH98,ML98,LHMZ98,HHZ00,AHZ01,JZ02,PZMZ02,CLH02,MHN02,VT04,GNB05,GNB05b,Z06,PTV07,CP08,S11b,%
VLSG17,VLSG17b,GSK18,TLLVPS19,GG20,HZ97,BPKZ07,KBPZ09,RA14,SKG10,SKS11,VSK14,VSK14b,VS15,KSG14,S18,MS19,MS19b,MS21a,MS21b}
The resulting granular-gas model can be referred to as the inelastic rough hard-sphere model (IRHSM). Needless to say, the IRHSM is even much more difficult to tackle with than the IHSM of frictionless, smooth particles. Therefore, in order to incorporate the influence of roughness and the associated rotational degrees of freedom on the dynamical properties and yet have a tractable model, it seems natural to construct an inelastic rough Maxwell model (IRMM), which would play a role with respect to the IRHSM similar to that played by the well-known IMM with respect to the IHSM (see Fig.~\ref{sketch}). To the best of our knowledge, such an IRMM has not been proposed or worked out before.

The aim of this paper is to construct a tractable IRMM and illustrate its main features by evaluating exactly all the collisional velocity moments of first and second degree, as well as the most relevant ones of
third and fourth degree. As expected, the results reduce to the known ones in the smooth limit (IMM)
\cite{EB02c,GS07,GS11,SG12}. We also display the results for the conservative case of elastic and perfectly rough particles ($\a=\b=1$), what constitutes the Maxwell analog of the Pidduck gas \cite{P22}. Moreover, we apply the exact knowledge of the isotropic second- and fourth-degree collisional moments to the study of the homogeneous cooling state (HCS), as described by the model.

The organization of the paper is as follows. In Sect.~\ref{sec2}, after summarizing the collision rules and the basic properties of the Boltzmann equation for a granular gas of inelastic and rough particles, and recalling the IRHSM, our IRMM is written down.  The core of the paper is presented in Sect.~\ref{sec3}, where the exact structure of the collisional moments is displayed in Table \ref{table1}, the explicit expressions of the coefficients in terms of inelasticity and roughness being moved to Appendix \ref{appB}, while some particularizations and consistency tests are shown in Appendix \ref{appC}. Next, Sect.~\ref{sec4} is devoted to the application of the results to the HCS. Finally, the paper is closed in Sect.~\ref{sec5} with some concluding remarks.

\section{Kinetic theory of inelastic and rough particles}\label{sec2}
We consider a granular gas made of inelastic and rough particles of diameter $\sigma$, mass $m$, and moment of inertial $I= (m\d^2/4)\kappa$, where the dimensionless moment of inertia ranges  from $\ka=0$ (mass concentrated at the center of the sphere) to a maximum value  $\ka=\frac{2}{3}$ (mass concentrated at the spherical surface), the value $\kappa=\frac{2}{5}$ corresponding to a uniform mass distribution.

\subsection{Collision rules}
\label{appA}

Let us denote by $\bc$ and $\bw$ the translational and angular velocities, respectively, and introduce the short-hand notation $\{\bc,\bw\}\to \bxi$ and $d\bc d\bw\to d\bxi$. A direct encounter between two particles $1$ and $2$ is characterized by
the precollisional velocities $(\bxi_1; \bxi_2)$,  the postcollisional
velocities $(\bxi_1'; \bxi_2')$, and  the collision unit vector $\bk=(\bx_2-\bx_1)/\sigma$   joining the centers of the two colliding particles.
The pre- and postcollisional velocities are related by
\begin{subequations}
\label{1,2}
\bal
\label{1}
 m \bc_1^\prime=m\bc_1 -\bJ,\quad &I \bw_1^\prime=I\bw_1 -{\d\over2}\,\bk\times\bJ,\\
 \label{2}
 m \bc_2^\prime=m\bc_2 +\bJ,\quad
 &I \bw_2^\prime=I\bw_2 -{\d\over2}\,\bk\times\bJ,
 \eal
\end{subequations}
where ${\bJ}$ is the impulse exerted by  particle $1$ on particle $2$. The  relative velocity of the points of the spheres at contact during a collision is
 $\bu=\bg-{\d\over2}\bk\times(\bw_1+\bw_2)$, where $\bg=\bc_1-\bc_2$ denotes the center-of-mass relative velocity; a similar relation holds for $\bu'$.

In the most basic model, an inelastic collision of two rough  particles is characterized by
 \be
\bk\cdot\bu^\prime=-\alpha(\bk\cdot\bu),\quad
\bk\times\bu^\prime=-\beta(\bk\times\bu),
 \ee{4}
where $0<\alpha\leq1$ and $-1\leq\beta\leq1$ are the coefficients of normal and tangential restitution, respectively. For an elastic collision of perfectly smooth particles one has $\alpha=1$ and $\beta=-1$, while $\alpha=1$ and $\beta=1$ for an elastic encounter of perfectly rough  particles.

It is possible to show that Eqs.~\eqref{4} imply \cite{G19,SKG10}
 \beq
\bJ={m\widetilde\alpha}(\bk\cdot\bg)\bk-m\widetilde\beta\bk\times\left(\bk\times\bg+\d{\bw_1+\bw_2\over 2}\right),
\label{6}
 \eeq
where
 \beq\label{7}
 \widetilde\alpha\equiv{1+\alpha\over2},\quad \widetilde\beta\equiv\frac{1+\beta}{2}\frac{\kappa}{1+\kappa}.
 \eeq
Equations \eqref{1,2} and (\ref{6})  express the  postcollisional velocities  in terms of the precollisional velocities and of the collision vector \cite{S10}.

For a restitution encounter, the pre- and postcollisional velocities are denoted by $(\bxi_1^*; \bxi_2^*)$ and $(\bxi_1; \bxi_2)$, respectively, and
the collision vector by $\bk^*=-\bk$.
One then has
 \be
\bk^*\cdot \bg=-\alpha\left(\bk^*\cdot\bg^*\right)=-\bk\cdot\bg,\quad  d\bxi_1^\ast  d\bxi_2^\ast ={1\over\alpha\beta^2}
d\bxi_1  d\bxi_2.
 \ee{14}

\subsection{Boltzmann equation}

Assuming molecular chaos, and in the absence of external forces or torques,  the Boltzmann equation for granular gases   is given by
  \beq
{\partial f \over \partial t}+\bc\cdot\bnabla f=J[\bxi\vert f,f],
\label{15a}
\eeq
where  $f(\bx,\bxi, t)$ is the one-particle VDF and $J[\bxi\vert f,f]$ is the bilinear collision operator. Quite generally, it can be written as
 \beq
J[\bxi_1\vert f,f]=\d^2\int d\bxi_2\int d\bk\,\left[\mathcal{F}(\bk^*\cdot\bg^*){f_1^\ast f_2^\ast\over\alpha\beta^2} -\mathcal{F}(\bk\cdot\bg)f_1f_2\right],
\label{15bb}
 \eeq
where use has been made of Eq.\ \eqref{14} and, as usual, the notation $f_1=f(\bxi_1)$, $f_2=f(\bxi_2)$, $f_1^*=f(\bxi_1^*)$, $f_2^*=f(\bxi_2^*)$ has been employed. Moreover, the function $\mathcal{F}(x)$ is proportional to the collision rate, its precise form defining the chosen kinetic model.

The so-called weak form of the Boltzmann equation is obtained by multiplying both sides of Eq.\ \eqref{15a} by an arbitrary trial function $\Psi(\bx,\bxi,t)$ and integrating over  $\bxi$. This yields
\beq
\label{16}
\partial_t \left(n\langle\Psi\rangle\right) +\bnabla\cdot \left(n \langle \bc\Psi\rangle\right)-n\langle (\partial_t+\bc\cdot\bnabla)\Psi\rangle=
n\mathcal{J}[\Psi],
\eeq
where
\beq
n(\bx,t)=\int d\bxi\, f(\bx,\bxi,t)
\label{n}
\eeq
is the local number density,
\beq
\langle\Psi\rangle ={1\over n(\bx,t)}\int d\bxi\, \Psi(\bx,\bxi,t)f(\bx,\bxi,t)
\label{avpsi}
\eeq
is the local average value of $\Psi$, and
\bal
\label{16J}
\mathcal{J}[\Psi]\equiv&\frac{1}{n}\int d\bxi\,\Psi(\bx,\bxi,t)J[\bxi\vert f,f]\nn
=& \frac{\d^2}{2n}\int d\bxi_1\int d\bxi_2\int d\bk\,\mathcal{F}(\bk\cdot\bg)f_1f_2 \left(\Psi_1'+\Psi_2'-\Psi_1-\Psi_2\right)
\eal
is the collisional production term of $\Psi$, with the conventional notation $\Psi_1=\Psi(\bxi_1)$, $\Psi_2=\Psi(\bxi_2)$, $\Psi_1'=\Psi(\bxi_1')$, $\Psi_2'=\Psi(\bxi_2')$.
In the second step  of Eq.\ \eqref{16J}, we have used the relationships (\ref{14}) and the standard symmetry properties of the collision term.

In particular, the flow velocity ($\mathbf{u}$), the mean angular velocity ($\bW$), and the granular temperatures ($T_t$, $T_r$, and $T$) correspond to $\Psi=\bc$, $\Psi=\bw$, $\Psi=V^2$, $\Psi=\omega^2$, and $\Psi=mV^2+I\omega^2$, respectively, where $\bV=\bc-\mathbf{u}$ denotes the peculiar velocity. Thus,
\beq
\mathbf{u}=\medio{\bc},\quad \bW=\medio{\bw},\quad T_t=\frac{m}{3}\medio{V^2},\quad T_r=\frac{I}{3}\medio{\omega^2},\quad T=\frac{1}{2}\left(T_t+T_r\right).
\eeq
The associated collisional production terms can be written as \cite{SKG10,KSG14}
\begin{subequations}
\label{10ab}
\beq
\mathcal{J}[\bc]=0,\quad \mathcal{J}[\bw]=-\zeta_\Omega \bW,\quad
 \mathcal{J}[V^2]=-\zeta_t\frac{3T_t}{m},\quad
 \mathcal{J}[\omega^2]=-\zeta_r\frac{3T_r}{I},
 \eeq
\beq
\label{10b}
\mathcal{J}[mV^2+I\omega^2]=-6\zeta T,\quad \zeta=\frac{T_t}{2T}\zeta_t+\frac{T_r}{2T}\zeta_r=\frac{\zeta_t+\theta\zeta_r}{1+\theta}.
\eeq
\end{subequations}
This defines the ``de-spinning'' rate coefficient $\zeta_\Omega$,  the energy production rates $\zeta_t$ and $\zeta_r$,  and the cooling rate  $\zeta$. Moreover, $\theta\equiv T_r/T_t$ is the rotational-to-translational temperature ratio.

\subsection{The inelastic rough hard-sphere model (IRHSM)}
If the gas is modeled by the IRHSM, one has $\mathcal{F}(x)=\Theta(x)x$, where $\Theta(x)$ is the Heaviside step function. In that case, the collision operator becomes
 \beq
J_{\text{HS}}[\bxi_1\mid f,f]=\d^2\int d\bxi_2\int_{+} d\bk\,(\bk\cdot\bg)\left({f_1^\ast f_2^\ast\over\alpha^2\beta^2} -f_1f_2\right),
\label{15b}
 \eeq
where the subscript ($+$) in the integration over $\bk$ denotes the constraint $\bk\cdot\bg>0$ and we have taken into account that $\Theta(\bk^*\cdot\bg^*)(\bk^*\cdot\bg^*)=\alpha^{-1}\Theta(\bk\cdot\bg)(\bk\cdot\bg)$. Analogously, Eq.\ \eqref{16J} becomes
\beq
\label{16HS}
\mathcal{J}_{\text{HS}}[\Psi]= \frac{\d^2}{2n}\int d\bxi_1\int d\bxi_2\int_+ d\bk\,(\bk\cdot\bg)f_1f_2 \left(\Psi_1'+\Psi_2'-\Psi_1-\Psi_2\right).
\eeq

If $\Psi(\bxi)$ is a polynomial and thus $\langle \Psi\rangle$ is a velocity moment, the collisional moment $\mathcal{J}_{\text{HS}}[\Psi]$ involves the full VDF or, equivalently, all the higher-degree moments. As a consequence, the infinite hierarchy of moments given by Eq.\ \eqref{16} cannot be solved, even in spatially uniform states, unless an approximate closure is applied.
For instance, if $f$ is approximated by a two-temperature Maxwellian, one finds \cite{HZ97,KSG14,G19}
\begin{subequations}
\label{13abcd}
\beq
\label{13a}
\zeta_\Omega^\hs\simeq\frac{5\nu_\hs}{3}\frac{\bt}{\ka},\quad \nu_\hs\equiv \frac{16}{5}n\sigma^2\sqrt{\pi T_t/m},
\eeq
\beq
\label{13b}
\zeta_t^\hs\simeq \frac{5\nu_\hs}{3}\left[\at(1-\at)+\bt(1-\bt)-\frac{\bt^2}{\ka}\theta(1+X)\right],\quad X\equiv \frac{I\Omega^2}{3T_r},
\eeq
\beq
\label{13c}
\zeta_r^\hs\simeq\frac{5\nu_\hs}{3}\frac{\bt}{\ka}\left[\left(1-\frac{\bt}{\ka}\right)\left(1+X\right)-\frac{\bt}{\theta}\right],
\eeq
\beq
\label{13d}
\zeta^\hs\simeq \frac{5}{12}\frac{\nu_\hs}{1+\theta}\left[1-\a^2+\frac{1-\b^2}{1+\ka}\theta\left(\frac{\ka}{\theta}+1+X\right)\right].
\eeq
\end{subequations}

\subsection{The inelastic rough Maxwell model (IRMM)}

The  problem mentioned below Eq.~\eqref{16HS} is also present in the conventional case of elastic collisions. However, if the collision rate is assumed to be a constant (Maxwell model), the collisional moments  involve moments of  a degree equal to or lower than the degree of the moment  $\langle \Psi\rangle$
\cite{TM80,E81,GS03,S09b,K10a}.

We now apply the same philosophy to a granular gas of inelastic and rough particles and propose the IRMM by choosing $\mathcal{F}(x)=\text{const}$ in Eq.\ \eqref{15bb}, i.e.,
 \beq
J_{\text{M}}[\bxi_1\vert f,f]=\frac{\nuM}{4\pi n}\int d\bxi_2\int d\bk\,\left({f_1^\ast f_2^\ast\over\alpha\beta^2} -f_1f_2\right),
\label{15M}
 \eeq
where $\nuM$  is an effective collision frequency.
Analogously, Eq.\ \eqref{16J} becomes
\beq
\label{16M}
\mathcal{J}_{\text{M}}[\Psi]= \frac{\nuM}{8\pi n^2}\int d\bxi_1\int d\bxi_2\int d\bk\,f_1f_2 \left(\Psi_1'+\Psi_2'-\Psi_1-\Psi_2\right).
\eeq

{\renewcommand{\arraystretch}{1.6}
\begin{table}
\begin{center}
\caption{Collisional moments according to the IRMM, Eqs.\  \eqref{15M} and \eqref{16M}.}\label{table1}
\begin{tabular}{@{}ll@{}}
\toprule
$\Psi_{k_1k_2}(\bxi)$&$-\nuM^{-1}\mathcal{J}_{\text{M}}[\Psi_{k_1k_2}]$\\
\midrule
$\bw$&$\varphi_{01\mid 01}\langle\bw\rangle$\\
$V^2$&$\chi_{20\mid 20}\langle V^2\rangle+\chi_{20\mid 02}\d^2\left(\langle \omega^2\rangle+\langle\bw\rangle^2\right)$\\
$\displaystyle{V_iV_j-\frac{V^2}{3}\delta_{ij}}$&$\displaystyle{\psi_{20\mid 20}\big(\langle V_iV_j\rangle-\frac{\medio{V^2}}{3}\delta_{ij}\big)
+\psi_{20\mid 02}\d^2\big(\langle\omega_i\omega_j\rangle+\langle\omega_i\rangle\langle\omega_j\rangle
-\frac{\medio{\omega^2}+\medio{\bw}^2}{3}\delta_{ij}\big)}
$\\
$\omega^2$&$\displaystyle{\chi_{02\mid 02}\left(\langle\omega^2\rangle+\langle\bw\rangle^2\right)+\frac{\chi_{02\mid 20}}{\d^2}\langle V^2\rangle}$\\
$\displaystyle{\omega_i\omega_j-\frac{\omega^2}{3}\delta_{ij}}$&$\displaystyle{\psi_{02\mid 02}\big(\langle\omega_i\omega_j\rangle+\langle\omega_i\rangle\langle\omega_j\rangle
-\frac{\medio{\omega^2}+\medio{\bw}^2}{3}\delta_{ij}\big)
+\frac{\psi_{02\mid 20}}{\d^2}\big(\langle V_iV_j\rangle-\frac{\medio{V^2}}{3}\delta_{ij}\big)}
$\\
$V_i\omega_j$&$\displaystyle{\psi_{11\mid 11}\langle V_i\omega_j\rangle}$\\
$V^2V_i$&$\displaystyle{\varphi_{30\mid 30}\langle V^2V_i\rangle
+\varphi_{30\mid 12}\d^2\left[2\langle \omega^2V_i\rangle-\langle(\bV\cdot\bw)\omega_i\rangle+4\langle\bw\rangle\cdot\langle\bw
 V_i\rangle\right.}$\\
 &$\quad\displaystyle{\left.-\langle\bw\rangle\cdot\langle\bV\omega_i\rangle-\langle \bV\cdot\bw\rangle\langle\omega_i\rangle\right]}$\\
 $\omega^2V_i$&$\displaystyle{\frac{\varphi_{12\mid 30}}{\d^2}\langle V^2V_i\rangle+\varphi_{12\mid 12}^{(1)}\langle\omega^2V_i\rangle+\varphi_{12\mid 12}^{(2)}\langle(\bV\cdot\bw)\omega_i\rangle
 }$\\
 &$\quad\displaystyle{+\varphi_{12\mid 12}^{(3)}\langle\bw\rangle\cdot\langle\bw V_i\rangle+\varphi_{12\mid 12}^{(4)}
 \left(\langle\bw\rangle\cdot\langle \bV\omega_i\rangle-\langle\bV\cdot\bw\rangle\langle\omega_i\rangle\right)}$\\
 $\left(\bV\cdot\bw\right)\omega_i$&$\displaystyle{\overline{\varphi}_{12\mid 12}^{(1)}
 \langle\left(\bV\cdot\bw\right)\omega_i\rangle
 +\overline{\varphi}_{12\mid 12}^{(2)}\langle\omega^2 V_i\rangle
  +\overline{\varphi}_{12\mid 12}^{(3)}\langle\bw\rangle\cdot\langle\bV\omega_i\rangle+\overline{\varphi}_{12\mid 12}^{(4)}\langle \bw\rangle\cdot\langle\bw V_i\rangle}$\\
&$\quad\displaystyle{ +\overline{\varphi}_{12\mid 12}^{(5)}\langle\bV\cdot\bw\rangle\langle\omega_i\rangle}$\\
$V^4$&$\displaystyle{\chi_{40\mid 40}^{(1)}\langle V^4\rangle+\chi_{40\mid 40}^{(2)}\langle V^2\rangle^2+\chi_{40\mid 40}^{(3)}\langle\bV\bV\rangle:\langle\bV\bV\rangle+\chi_{40\mid 04}\d^4\big(\langle\omega^4\rangle+\langle\omega^2\rangle^2}$\\
&$\quad\displaystyle{+2\langle\bw\bw\rangle:\langle\bw\bw\rangle+4\langle\omega^2\bw\rangle\cdot\langle\bw\rangle\big)+\chi_{40\mid 22}^{(1)}\d^2\big[2\langle V^2\omega^2\rangle+2\langle V^2\rangle\langle\omega^2\rangle}$\\
&$\quad\displaystyle{-\langle\left(\bV\cdot\bw\right)^2\rangle-\langle\bV\bV\rangle:\langle\bw\bw\rangle
-2\langle\left(\bV\cdot\bw\right)\bV\rangle\cdot\langle\bw\rangle+4\langle V^2\bw\rangle\cdot\langle\bw\rangle\big]}$\\
&$\quad\displaystyle{+\chi_{40\mid 22}^{(2)}\d^2\left(\langle\bV\cdot\bw\rangle^2+\langle \bV\bw\rangle:\langle \bV\bw\rangle
-4\langle \bV\bw\rangle:\langle \bw\bV\rangle\right)}$\\
$\omega^4$&$\displaystyle{\frac{\chi_{04\mid 40}}{\d^4}\left(\medio{V^4}+\medio{V^2}^2+2\medio{\bV\bV}:\medio{\bV\bV}\right)+\frac{\chi_{04\mid 22}^{(1)}}{\d^2}\big[2\medio{V^2\omega^2}+2\medio{V^2}\medio{\omega^2}}$\\
&$\quad\displaystyle{-\medio{\left(\bV\cdot\bw\right)^2}-\medio{\bV\bV}:\medio{\bw\bw}\big]+\frac{\chi_{04\mid 22}^{(2)}}{\d^2}\big[2\medio{\left(\bV\cdot\bw\right)\bV}\cdot\medio{\bw}}$\\
&$\quad\displaystyle{
-4\medio{V^2\bw}\cdot\medio{\bw}-\medio{\bV\cdot\bw}^2-
\medio{\bV\bw}:\medio{\bV\bw}+4\medio{\bV\bw}:\medio{\bw\bV}\big]}$\\
&$\quad\displaystyle{+\chi_{04\mid 04}^{(1)}\medio{\omega^4}+\chi_{04\mid 04}^{(2)}\medio{\omega^2}^2+\chi_{04\mid 04}^{(3)}\medio{\bw\bw}:\medio{\bw\bw}+\chi_{04\mid 04}^{(4)}\medio{\omega^2\bw}\cdot\medio{\bw}}$\\
$V^2\omega^2$&$\displaystyle{\frac{\chi_{22\mid 40}^{(1)}}{\d^2}\left(\medio{V^4}+\medio{V^2}^2\right)+\frac{\chi_{22\mid 40}^{(2)}}{\d^2}\medio{\bV\bV}:\medio{\bV\bV}+\chi_{22\mid 22}^{(1)}\medio{V^2\omega^2}}$\\
&$\quad \displaystyle{+\chi_{22\mid 22}^{(2)}\medio{V^2}\medio{\omega^2}+\chi_{22\mid 22}^{(3)}\medio{\left(\bV\cdot\bw\right)^2}}+\chi_{22\mid 22}^{(4)}\medio{\bV\bV}:\medio{\bw\bw}$\\
&$\quad\displaystyle{+\chi_{22\mid 22}^{(5)}\medio{\left(\bV\cdot\bw\right)\bV}\cdot\medio{\bw}+\chi_{22\mid 22}^{(6)}\medio{V^2\bw}\cdot\medio{\bw}+\chi_{22\mid 22}^{(7)}\medio{\bV\cdot\bw}^2}$\\
&$\quad\displaystyle{+\chi_{22\mid 22}^{(8)}\medio{\bV\bw}:\medio{\bV\bw}+\chi_{22\mid 22}^{(9)}\medio{\bV\bw}:\medio{\bw\bV}+\chi_{22\mid 04}^{(1)}\d^2\left(\medio{\omega^4}+\medio{\omega^2}^2\right)}$\\
&$\quad\displaystyle{\chi_{22\mid 04}^{(2)}\d^2\medio{\bw\bw}:\medio{\bw\bw}+\chi_{22\mid 04}^{(3)}\d^2\medio{\omega^2\bw}\cdot\medio{\bw}}$\\
$\left(\bV\cdot\bw\right)^2$&$\displaystyle{\frac{\overline{\chi}_{22\mid 40}}{\d^2}\left(\medio{V^2}^2-\medio{\bV\bV}:\medio{\bV\bV}\right)+\overline{\chi}_{22\mid 22}^{(1)}\medio{\left(\bV\cdot\bw\right)^2}+\overline{\chi}_{22\mid 22}^{(2)}\medio{V^2\omega^2}}$\\
&$\quad\displaystyle{+\overline{\chi}_{22\mid 22}^{(3)}\medio{V^2}\medio{\omega^2}+\overline{\chi}_{22\mid 22}^{(4)}\medio{\bV\bV}:\medio{\bw\bw}+\overline{\chi}_{22\mid 22}^{(5)}\medio{\left(\bV\cdot\bw\right)\bV}\cdot\medio{\bw}}$\\
&$\quad\displaystyle{+\overline{\chi}_{22\mid 22}^{(6)}\medio{V^2\bw}\cdot\medio{\bw}+\overline{\chi}_{22\mid 22}^{(7)}\medio{\bV\cdot\bw}^2+\overline{\chi}_{22\mid 22}^{(8)}\medio{\bV\bw}:\medio{\bV\bw}}$\\
&$\quad\displaystyle{+\overline{\chi}_{22\mid 22}^{(9)}\medio{\bV\bw}:\medio{\bw\bV}+\overline{\chi}_{22\mid 04}\d^2\left(\medio{\omega^2}^2-\medio{\bw\bw}:\medio{\bw\bw}\right)}$\\
\botrule
\end{tabular}
\end{center}
\end{table}
}

{\renewcommand{\arraystretch}{1.8}
\begin{table}
\begin{center}
\caption{Comparison between the basic production rates in the IRHSM and IRMM with two main choices of $\nu_{\text{M}}$.}\label{table5}
\begin{tabular}{@{}cccc@{}}
\toprule
Choice for $\nu_{\text{M}}$&$\hs\simeq\text{M}$&$\alpha$&$\beta$\\
\midrule
$\frac{5}{4}\nu_\hs$&$\zeta_\Omega^\hs\simeq\zeta_\Omega^{\text{M}}$&Arbitrary&Arbitrary\\
&$\zeta_r^\hs\simeq\zeta_r^{\text{M}}$&Arbitrary&Arbitrary\\
&$\zeta_t^\hs\simeq\zeta_t^{\text{M}}$&$\alpha_{\text{M}}\leftrightarrow \alpha_{\text{HS}}$&Arbitrary\\
&$\zeta^\hs\simeq\zeta^{\text{M}}$&$\alpha_{\text{M}}\leftrightarrow \alpha_{\text{HS}}$&Arbitrary\\
$\frac{5}{2}\nu_\hs$&$\zeta_t^\hs\simeq\zeta_t^{\text{M}}$&Arbitrary&$-1$\\
&$\zeta^\hs\simeq\zeta^{\text{M}}$&Arbitrary&$\pm1$\\
\botrule
\end{tabular}
\end{center}
\end{table}
}

\section{Collisional moments}
\label{sec3}

Using the collision rules \eqref{1,2} and \eqref{6} in Eq.\ \eqref{16M}, it is possible to evaluate  the collisional moments in terms of velocity moments. After some tedious work, we obtained all the collisional moments of first and second degree, plus the most relevant ones  of third and fourth degree. Their structure is shown in Table \ref{table1}, where the expected result $\mathcal{J}_{\text{M}}[\bV]=0$ (momentum conservation) is not included. This yields a total number of $13$ independent collisional moments and $60$ dimensionless coefficients.

We have adopted the following criterion for the notation of the coefficients. Let us first denote by $\Psi_{k_1k_2}(\bxi)=\Psi_{k_1}^{(1)}(\bV)\Psi_{k_2}^{(2)}(\bw)$  a homogeneous velocity polynomial  of degrees $k_1$ and $k_2$ with respect to $\bV$ and $\bw$, respectively, i.e., $\Psi_{k_1}^{(1)}(\lambda \bV)=\lambda^{k_1}\Psi_{k_1}^{(1)}(\bV)$ and $\Psi_{k_2}^{(2)}(\lambda \bw)=\lambda^{k_2}\Psi_{k_2}^{(2)}(\bw)$. Then, a coefficient of the form $Y_{k_1 k_2\mid \ell_1\ell_2}$  (with $k_1+k_2=\ell_1+\ell_2$) corresponds to the collisional moment $\mathcal{J}_{\text{M}}[\Psi_{k_1k_2}]$  and  accompanies a product of the form $\langle \Psi_{i_1i_2}(\bxi)\rangle\langle \Psi_{j_1j_2}(\bxi)\rangle$ with $i_1+j_1=\ell_1$ and $i_2+j_2=\ell_2$. When, for a given $\mathcal{J}_{\text{M}}[\Psi_{k_1k_2}]$, more than a coefficient of the form $Y_{k_1 k_2\mid \ell_1\ell_2}$ exists, each one of them is distinguished by a superscript. A coefficient of the form $Y_{k_1 k_2\mid k_1k_2}$ or $Y_{k_1 k_2\mid k_1k_2}^{(1)}$ is a \emph{diagonal} one because it couples $\mathcal{J}_{\text{M}}[\Psi_{k_1k_2}]$ to $\langle \Psi_{k_1k_2}(\bxi)\rangle$.
Finally, we use the Greek letters $Y=\chi$, $Y=\varphi$, and $Y=\psi$ for the coefficients in $\mathcal{J}_{\text{M}}[\Psi_{k_1k_2}]$ associated with scalar, vector, and tensor quantities $\Psi_{k_1k_2}(\bxi)$, respectively; moreover, an overline is used if the definition of $\Psi_{k_1k_2}(\bxi)$ contains the inner product $\bV\cdot\bw$.

It must be remarked that the results for $\Psi_{01}(\bxi)=\bw$, $\Psi_{20}(\bxi)=V^2$, and $\Psi_{02}(\bxi)=\omega^2$ are also valid in the case of a hybrid model where $\mathcal{F}(x)=\text{const}\times\Theta(x)$. However, this does not happen in the case of the other quantities in Table \ref{table1}, that is, the associated collisional moments cannot be expressed only in terms of velocity moments of the same or lower degree if $\mathcal{F}(x)=\text{const}\times\Theta(x)$.

The exact explicit expressions for the $60$ coefficients appearing in Table \ref{table1}  in terms of $\at$, $\bt$, and $\ka$ are given in Appendix \ref{appB}.
The particularization of the coefficients to the IMM ($\a<1$, $\b=-1$) and to the Pidduck model ($\a=\b=1$) \cite{P22} is presented in Appendix \ref{appC}, where some consistency tests  are performed.

The basic production rates are those defined by Eqs.\ \eqref{10ab}. According to Table \ref{table1},
\begin{subequations}
\label{16abcd}
\beq
\label{16a}
\zeta_\Omega^{\text{M}}=\nu_{\text{M}}\varphi_{01\mid 01}
=\frac{4\nu_{\text{M}}}{3}\frac{\bt}{\ka},
\eeq
\bal
\label{16b}
\zeta_t^{\text{M}}=&\nu_{\text{M}}\left[\chi_{20\mid 20}+\frac{4\theta}{\ka}\chi_{20\mid 02}\left(1+X\right)\right]\nn
 =&\frac{2\nu_{\text{M}}}{3}\left[\at(1-\at)+2\bt(1-\bt)-\frac{2\bt^2}{\ka}\theta(1+X)\right],
\eal
\beq
\label{16c}
\zeta_r^{\text{M}}=\nu_{\text{M}}\left[\frac{\ka}{4\theta}\chi_{02\mid 20}+\chi_{02\mid 02}\left(1+X\right)\right]=\frac{4\nu_{\text{M}}}{3}\frac{\bt}{\ka}\left[\left(1-\frac{\bt}{\ka}\right)\left(1+X\right)-\frac{\bt}{\theta}\right],
\eeq
\bal
\label{16d}
\zeta^{\text{M}}=&\frac{\nu_{\text{M}}}{1+\theta}\left[\chi_{20\mid 20}+\frac{\ka}{4}\chi_{02\mid 20}+\left(\frac{4}{\ka}\chi_{20\mid 02}+\chi_{02\mid 02}\right)\theta\left(1+X\right)\right]\nn
=& \frac{1}{6}\frac{\nu_{\text{M}}}{1+\theta}\left[1-\a^2+2\frac{1-\b^2}{1+\ka}\theta\left(\frac{\ka}{\theta}+1+X\right)\right],
\eal
\end{subequations}
where in the second equalities we have made use of Eqs.\ \eqref{B1}, \eqref{B2}, and \eqref{B4}.

Comparison with the approximate results for the IRHSM, Eqs.\ \eqref{13abcd}, shows that the choices $\nu_{\text{M}}=\frac{5}{4}\nu_\hs$ and $\nu_{\text{M}}=\frac{5}{2}\nu_\hs$ are directly related to inelasticity and roughness, respectively. Thus,  in comparison with the IRHSM, the IRMM lessens the impact of inelasticity on energy dissipation, relative to the impact of roughness. As a consequence, there is no a unique choice for $\nu_{\text{M}}$ allowing for an agreement between the  IRHSM and IRMM basic production rates for arbitrary $\alpha$ and $\beta$. This is reminiscent of the inability of the Bhatnagar--Gross--Krook (BGK) kinetic model to reproduce the Boltzmann shear viscosity and thermal conductivity with a single collision frequency \cite{C88}.

A way of circumventing the impossibility of matching Eqs.~\eqref{13abcd} and \eqref{16abcd} with a unique relationship between $\nu_\hs$ and $\nu_{\text{M}}$ consists of choosing $\nu_{\text{M}}=\frac{5}{4}\nu_\hs$ and then assuming the following mapping between the coefficients of normal restitution in the IRHSM ($\alpha_{\text{HS}}$) and the IRMM ($\alpha_{\text{M}}$): $1-\alpha_{\text{M}}^2=2(1-\alpha_{\text{HS}}^2)$. For instance, $\alpha_{\text{M}}=0.4$, $0.6$, $0.8$, and $1$ would correspond to $\alpha_{\text{HS}}=0.76$, $0.82$, $0.91$, and $1$, respectively. Table \ref{table5} summarizes the consequences of the two main choices $\nu_{\text{M}}=\frac{5}{4}\nu_\hs$ and $\nu_{\text{M}}=\frac{5}{2}\nu_\hs$.

Alternatively, and in analogy with a BGK-like model for the IRHSM \cite{S11a}, one can replace the IRMM collision operator  \eqref{15M} by
\begin{equation}
\label{17}
J_{\text{M}}[\bxi\mid f,f]\to J_{\text{M}}[\bxi\mid f,f]+\nu_{\text{M}}\gamma\frac{\partial}{\partial\bV}\cdot\left(\bV f\right),\quad \gamma\equiv \frac{1-\alpha^2}{12}.
\end{equation}
This modified IRMM keeps being amenable to an exact evaluation of the associated collisional moments and, in addition, allows one to recover Eqs.~\eqref{13abcd} if $\nu_{\text{M}}=\frac{5}{4}\nu_\hs$.
More specifically, only the {diagonal} coefficients are affected by the modification \eqref{17}: $Y_{k_1k_2\mid k_1k_2}\to Y_{k_1k_2\mid k_1k_2}+k_1\gamma$ and $Y_{k_1k_2\mid k_1k_2}^{(1)}\to Y_{k_1k_2\mid k_1k_2}^{(1)}+k_1\gamma$.

On the other hand, in this paper we restrict ourselves to the IRMM as defined by Eq.~\eqref{15M}, that is, without the extra term appearing in Eq.~\eqref{17}.
The reason is that we regard the IRMM as a mathematical model on its own and not necessarily as a model intended to mimic the properties of the IRHSM, for which alternative approximate tools are already available
\cite{HZ97,BPKZ07,SKG10,SKS11,VSK14,KSG14,S18,MS19,MS19b,G19,MS21a,MS21b}.

\section{Application to the Homogeneous Cooling State (HCS)}
\label{sec4}

In the absence of gradients or any external driving, Eq.\ \eqref{16} becomes
\beq
\frac{d}{dt}\medio{\Psi}=\mathcal{J}[\Psi].
\eeq
For sufficiently long times, the system asymptotically reaches the HCS, which is characterized by a uniform, isotropic, and stationary \emph{scaled} VDF \cite{G19,KSG14}
\beq
\phi(\bcc,\bww)=\frac{\sqrt{4T_tT_r/mI}}{n}f(\bV,\bw;t),\quad \bcc\equiv\frac{\bV}{\sqrt{2T_t(t)/m}},\quad \bww\equiv\frac{\bw}{\sqrt{2T_r(t)/m}}.
\eeq
In particular, $X\to 0$ and $\dot{\theta}\to 0$. The latter condition implies $\zeta_t-\zeta_r\to 0$, which yields the HCS temperature ratio
\begin{subequations}
\label{22ab}
\beq
\theta=h+\sqrt{1+h^2},
\eeq
\begin{equation}
h\equiv\frac{\ka}{8}\frac{\chi_{02\mid 02}-\chi_{20\mid 20}}{\chi_{20\mid 02}}=\frac{1+\ka}{2\ka(1+\b)}\left[c\frac{1+\ka}{2}\frac{1-\a^2}{1+\b}-(1-\ka)(1-\b)\right],
\end{equation}
\end{subequations}
with $c=1$. As expected from the discussion in Sect.~\ref{sec3}, in the case of the IRHSM, the two-temperature Maxwellian approximation
[see Eqs.~\eqref{13abcd}] also yields Eqs.~\eqref{22ab}, except that $c=2$ \cite{KSG14,G19}.
\begin{figure}[h]
\centering
\includegraphics[width=0.5\textwidth]{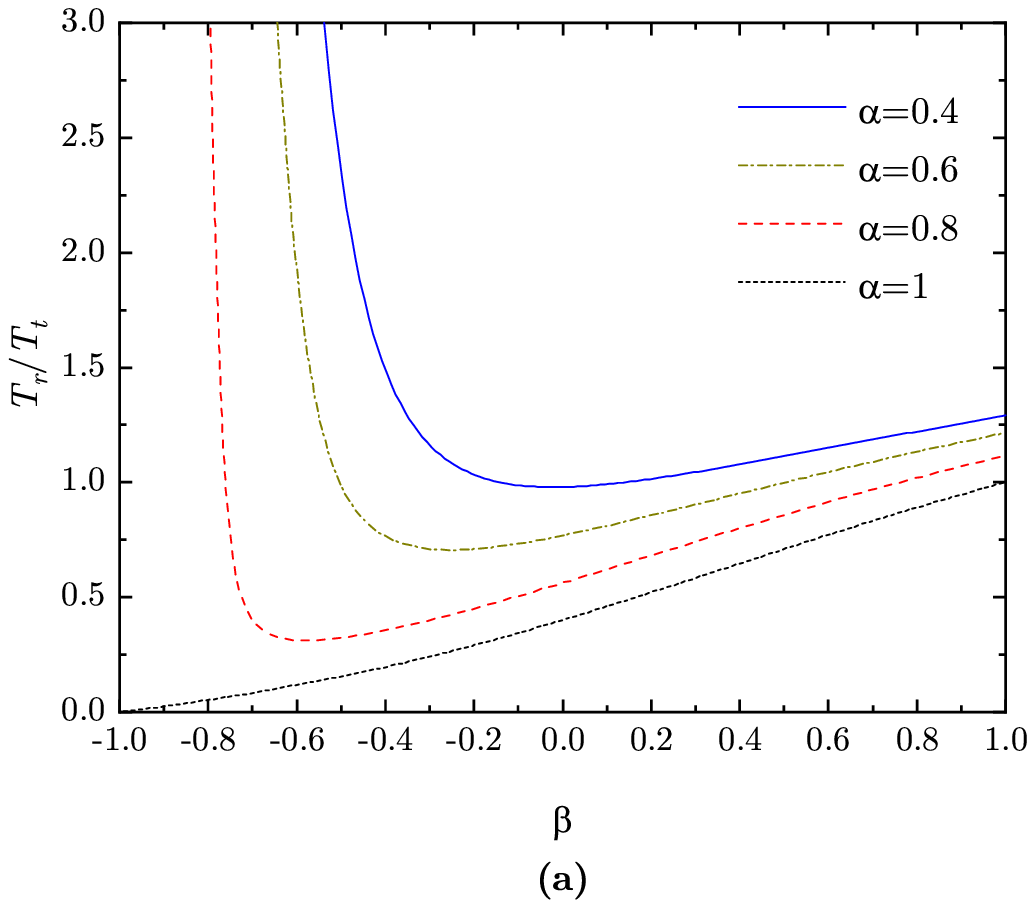}\includegraphics[width=0.5\textwidth]{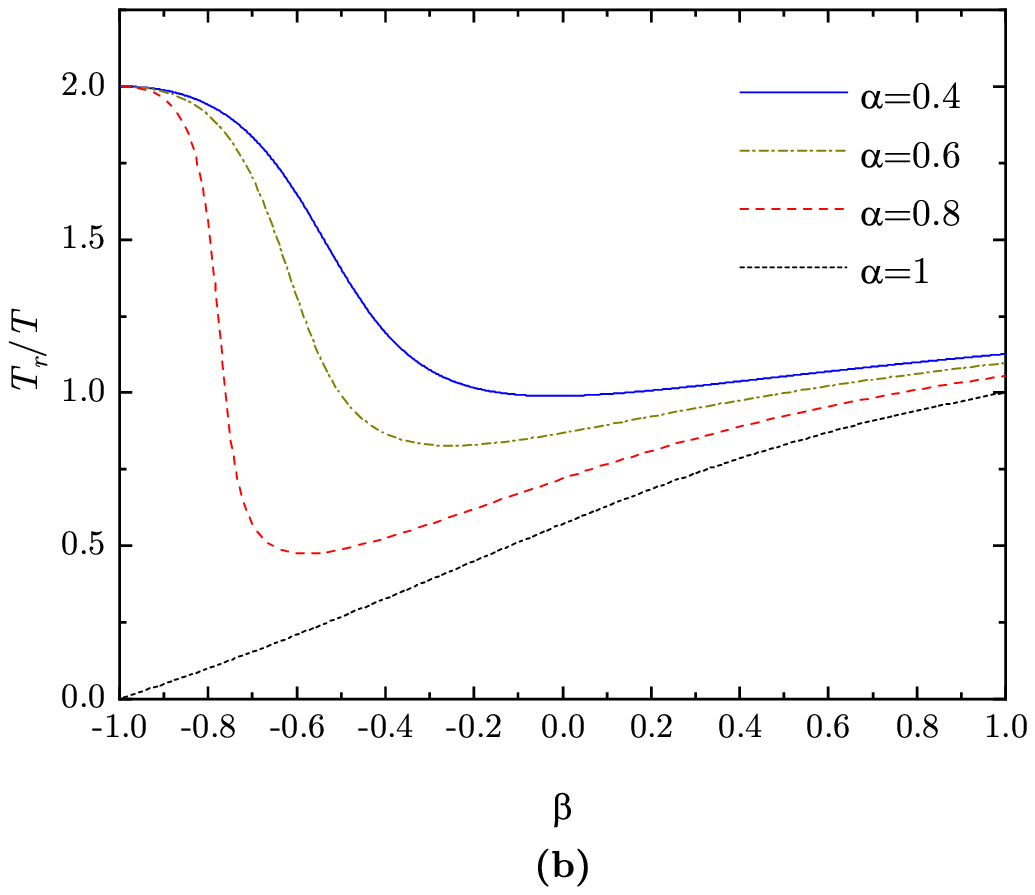}\\
\caption{Plot of the temperature ratios \textbf{a} $T_r/T_t=\theta$ and \textbf{b} $T_r/T=2\theta/(1+\theta)$ as functions of $\b$ for $\kappa=\frac{2}{5}$ and $\alpha= 0.4$ (---), $0.6$ (--$\cdot$--), $0.8$ (-- --), and $1$ ($\cdots$).}\label{fig1}
\end{figure}

Figure \ref{fig1} shows the $\beta$-dependence of the temperature ratios $T_r/T_t$ and  $T_r/T$ for uniform particles ($\kappa=\frac{2}{5}$) and several representative values of the coefficient of normal restitution. We observe that, except in the case of elastic collisions ($\a=1$), $T_r/T_t\to\infty$ and, hence, $T_r/T\to2$ in the quasi-smooth limit ($\b\to -1$).

The scaled fourth-degree moments are $\medio{c^4}=\frac{9}{4} \medio{V^4}/\medio{V^2}^2$, $\medio{w^4}=\frac{9}{4}\medio{\omega^4}/\medio{\omega^2}^2$, $\medio{c^2w^2}=\frac{9}{4}\medio{V^2\omega^2}/\medio{V^2}\medio{\omega^2}$, and $\medio{(\bcc\cdot\bww)^2}=\frac{9}{4}\medio{(\bV\cdot\bw)^2}/\medio{V^2}\medio{\omega^2}$. Assuming the stationary  VDF $\phi(\bcc,\bww)$ has not been reached yet, the evolution equations for those moments are
\begin{subequations}
\label{23abcd}
\bal
\frac{d\medio{c^4}}{\nuM dt}=&\frac{1}{\medio{V^2}^2}\frac{\mathcal{J}_{\text{M}}[V^4]}{\nu_{\text{M}}}-2\frac{\medio{V^4}}{\medio{V^2}^3}
\frac{\mathcal{J}_{\text{M}}[V^2]}{\nu_{\text{M}}}\nn
=&-\chi_{40\mid 40}^{(1)}\frac{\medio{V^4}}{\medio{V^2}^2}-\chi_{40\mid 40}^{(2)}-\frac{1}{3}\chi_{40\mid 40}^{(3)}-\frac{16\theta^2}{\ka^2}\chi_{40\mid 04}\left(\frac{\medio{\omega^4}}{\medio{\omega^2}^2}+\frac{5}{3}\right)-\frac{4\theta}{\ka}\chi_{40\mid 22}^{(1)}\nn
&\times\left[2\frac{\medio{V^2\omega^2}}{\medio{V^2}\medio{\omega^2}}-\frac{\medio{(\bV\cdot\bw)^2}}{\medio{V^2}\medio{\omega^2}}+\frac{5}{3}\right]
+2\left(\chi_{20\mid 20 }+\frac{4\theta}{\ka}\chi_{20\mid 02 }\right)\frac{\medio{V^4}}{\medio{V^2}^2},
\eal
\bal
\frac{d\medio{w^4}}{\nuM dt}=&\frac{1}{\medio{\omega^2}^2}\frac{\mathcal{J}_{\text{M}}[\omega^4]}{\nu_{\text{M}}}-2\frac{\medio{\omega^4}}{\medio{\omega^2}^3}
\frac{\mathcal{J}_{\text{M}}[\omega^2]}{\nu_{\text{M}}}\nn
=&-\frac{\ka^2}{16\theta^2}\chi_{04\mid 40}\left(\frac{\medio{V^4}}{\medio{V^2}^2}+\frac{5}{3}\right)-\frac{\ka}{4\theta}\chi_{04\mid 22 }^{(1)}\left[2\frac{\medio{V^2\omega^2}}{\medio{V^2}\medio{\omega^2}}-\frac{\medio{(\bV\cdot\bw)^2}}{\medio{V^2}\medio{\omega^2}}+\frac{5}{3}\right]\nn
&-\chi_{04\mid 04}^{(1)}\frac{\medio{\omega^4}}{\medio{\omega^2}^2}-\chi_{04\mid 04}^{(2)}-\frac{1}{3}\chi_{04\mid 04}^{(3)}
+2\left(\chi_{02\mid 02 }+\frac{\ka}{4\theta}\chi_{02\mid 20 }\right)\frac{\medio{\omega^4}}{\medio{\omega^2}^2},
\eal
\bal
\frac{d\medio{c^2w^2}}{\nuM dt}=&\frac{1}{\medio{V^2}\medio{\omega^2}}\frac{\mathcal{J}_{\text{M}}[V^2\omega^2]}{\nu_{\text{M}}}-\frac{\medio{V^2\omega^2}}{\medio{V^2}^2
\medio{\omega^2}}\frac{\mathcal{J}_{\text{M}}[V^2]}{\nu_{\text{M}}}-\frac{\medio{V^2\omega^2}}{\medio{V^2}
\medio{\omega^2}^2}\frac{\mathcal{J}_{\text{M}}[\omega^2]}{\nu_{\text{M}}}\nn
=&-\frac{\ka}{4\theta}\chi_{22\mid 40}^{(1)}\left(\frac{\medio{V^4}}{\medio{V^2}^2}+1\right)-\frac{\ka}{12\theta}\chi_{22\mid 40}^{(2)}
-\chi_{22\mid 22}^{(1)}\frac{\medio{V^2\omega^2}}{\medio{V^2}\medio{\omega^2}}-\chi_{22\mid 22}^{(2)}\nn
&-\chi_{22\mid 22}^{(3)}\frac{\medio{(\bV\cdot\bw)^2}}{\medio{V^2}\medio{\omega^2}}-\frac{1}{3}\chi_{22\mid 22}^{(4)}-\frac{4\theta}{\ka}\chi_{22\mid 04}^{(1)}\left(\frac{\medio{\omega^4}}{\medio{\omega^2}^2}+1\right)-\frac{4\theta}{3\ka}\chi_{22\mid 04}^{(2)}\nn
&+\left(\chi_{20\mid 20}+\frac{4\theta}{\ka}\chi_{20\mid 02}+\chi_{02\mid 02}+\frac{\ka}{4\theta}\chi_{02\mid 20}\right)\frac{\medio{V^2\omega^2}}{\medio{V^2}\medio{\omega^2}},
\eal
\bal
\frac{d\medio{(\bcc\cdot\bww)^2}}{\nuM dt}=&\frac{1}{\medio{V^2}\medio{\omega^2}}\frac{\mathcal{J}_{\text{M}}[(\bV\cdot\bw)^2]}{\nu_{\text{M}}}-\frac{\medio{(\bV\cdot\bw)^2}}{\medio{V^2}^2
\medio{\omega^2}}\frac{\mathcal{J}_{\text{M}}[V^2]}{\nu_{\text{M}}}-\frac{\medio{(\bV\cdot\bw)^2}}{\medio{V^2}
\medio{\omega^2}^2}\frac{\mathcal{J}_{\text{M}}[\omega^2]}{\nu_{\text{M}}}\nn
=&-\frac{\ka}{6\theta}\overline{\chi}_{22\mid 40}-\overline{\chi}_{22\mid 22}^{(1)}\frac{\medio{(\bV\cdot\bw)^2}}{\medio{V^2}\medio{\omega^2}}-\overline{\chi}_{22\mid 22}^{(2)}\frac{\medio{V^2\omega^2}}{\medio{V^2}\medio{\omega^2}}-\overline{\chi}_{22\mid 22}^{(3)}-\frac{1}{3}\overline{\chi}_{22\mid 22}^{(4)}\nn
&-\frac{8\theta}{3\ka}\overline{\chi}_{22\mid 04}
+\left(\chi_{20\mid 20}+\frac{4\theta}{\ka}\chi_{20\mid 02}+\chi_{02\mid 02}+\frac{\ka}{4\theta}\chi_{02\mid 20}\right)\frac{\medio{(\bV\cdot\bw)^2}}{\medio{V^2}\medio{\omega^2}}.
\eal
\end{subequations}

The departure of the HCS scaled  VDF $\phi(\bcc,\bww)$ from the Maxwellian $\pi^{-3}e^{-c^2-w^2}$ can be measured by the four cumulants \cite{SKS11,VSK14,VSK14b}
\begin{subequations}
\label{cumu}
\beq
a_{20}^{(0)}\equiv\frac{3}{5}\frac{\medio{V^4}}{\medio{V^2}^2}-1=\frac{4}{15}\medio{c^4}-1,\quad a_{02}^{(0)}\equiv\frac{3}{5}\frac{\medio{\omega^4}}{\medio{\omega^2}^2}-1=\frac{4}{15}\medio{w^4}-1,
\eeq
\beq
 a_{11}^{(0)}\equiv\frac{\medio{V^2\omega^2}}{\medio{V^2}\medio{\omega^2}}-1=\frac{4}{9}\medio{c^2w^2}-1,
 \eeq
\beq
a_{00}^{(1)}\equiv\frac{6}{5}\frac{\medio{(\bV\cdot\bw)^2}-\frac{1}{3}\medio{V^2\omega^2}}{\medio{V^2}\medio{\omega^2}}=
\frac{8}{15}\left[\medio{(\bcc\cdot\bww)^2}
-\frac{1}{3}\medio{c^2w^2}\right].
\eeq
\end{subequations}
In terms of those cumulants, Eqs.\ \eqref{23abcd} can be recast as
\beq
\label{25}
\nuM^{-1}\frac{d}{ dt}\left[
\begin{array}{c}
a_{20}^{(0)}\\
a_{02}^{(0)}\\
a_{11}^{(0)}\\
a_{00}^{(1)}
\end{array}
\right]+
\mathsf{M}\cdot \left[
\begin{array}{c}
a_{20}^{(0)}\\
a_{02}^{(0)}\\
a_{11}^{(0)}\\
a_{00}^{(1)}
\end{array}
\right]
=-\mathsf{L},
\eeq
where the elements of the matrices $\mathsf{M}$ and $\mathsf{L}$ are
\begin{subequations}
\beq
M_{11}=\chi_{40\mid 40}^{(1)}-2\chi_{20\mid 20}-\frac{8\theta}{\ka}\chi_{20\mid 02},
\eeq
\beq
M_{12}=\frac{16\theta^2}{\ka^2}\chi_{40\mid 04},\quad
M_{13}=\frac{4\theta}{\ka}\chi_{40\mid 22}^{(1)},\quad
M_{14}=-\frac{2\theta}{\ka}\chi_{40\mid 22}^{(1)},
\eeq
\beq
M_{21}=\frac{\ka^2}{16\theta^2}\chi_{04\mid 40},\quad M_{22}=\chi_{04\mid 04}^{(1)}-2\chi_{02\mid 02}-\frac{\ka}{2\theta}\chi_{02\mid 20},
\eeq
\beq
M_{23}=\frac{\ka}{4\theta}\chi_{04\mid 22}^{(1)},\quad M_{24}=-\frac{\ka}{8\theta}\chi_{04\mid 22}^{(1)},
\eeq
\beq
M_{31}=\frac{\ka}{4\theta}\chi_{22\mid 40}^{(1)},\quad M_{32}=\frac{4\theta}{\ka}\chi_{22\mid 04}^{(1)},\quad M_{34}=\frac{1}{2}\chi_{22\mid 22}^{(3)},
\eeq
\beq
M_{33}=\frac{3}{5}\left[\chi_{22\mid 22}^{(1)}-\chi_{20\mid 20}-\frac{4\theta}{\ka}\chi_{20\mid 02}-\chi_{02\mid 02}-\frac{\ka}{4\theta}\chi_{02\mid 20}\right]+\frac{1}{5}\chi_{22\mid 22}^{(3)},
\eeq
\beq
M_{41}=0,\quad M_{42}=0,\quad M_{43}=\overline{\chi}_{22\mid 22}^{(2)}+\frac{2}{5}M_{44},
\eeq
\beq
M_{44}=\frac{5}{6}\left[\overline{\chi}_{22\mid 22}^{(1)}-\chi_{20\mid 20}-\frac{4\theta}{\ka}\chi_{20\mid 02}-\chi_{02\mid 02}-\frac{\ka}{4\theta}\chi_{02\mid 20}\right],
\eeq
\end{subequations}
\begin{subequations}
\beq
L_1=M_{11}+2M_{12}+2M_{13}+\frac{3}{5}\chi_{40\mid 40}^{(2)}+\frac{1}{5}\chi_{40\mid 40}^{(3)}
,
\eeq
\beq
L_2=2M_{21}+M_{22}+2M_{23}+\frac{3}{5}\chi_{04\mid 04}^{(2)}+\frac{1}{5}\chi_{04\mid 04}^{(3)},
\eeq
\beq
L_3=\frac{8}{5}M_{31}+\frac{8}{5}M_{32}+M_{33}+\frac{\ka}{20\theta}\chi_{22\mid 40}^{(2)}+\frac{3}{5}\chi_{22\mid 22}^{(2)}+\frac{1}{5}\chi_{22\mid 22}^{(4)}+\frac{4\theta}{5\ka}\chi_{22\mid 04}^{(2)},
\eeq
\beq
 L_4=M_{43}+\frac{\ka}{6\theta}\overline{\chi}_{22\mid 40}+\overline{\chi}_{22\mid 22}^{(3)}+\frac{1}{3}\overline{\chi}_{22\mid 22}^{(4)}+\frac{8\theta}{3\ka}\overline{\chi}_{22\mid 04}.
 \eeq
\end{subequations}

\begin{figure}[h]
\centering
\includegraphics[width=0.5\textwidth]{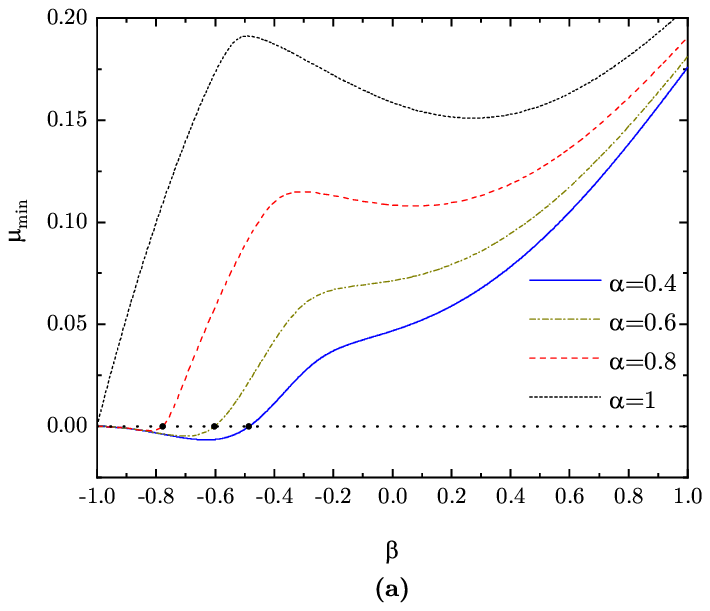}\includegraphics[width=0.5\textwidth]{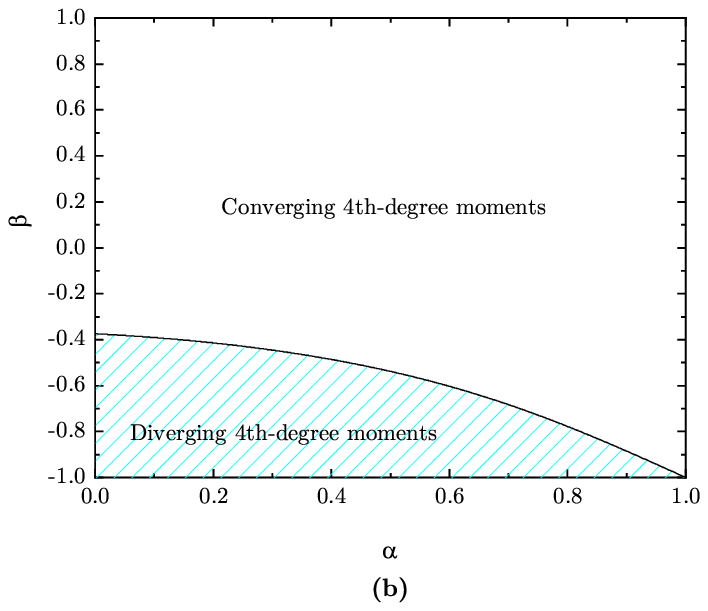}
\caption{\textbf{a} Plot of  the smallest eigenvalue $\mu_{\min}$ as a function  of $\b$ for $\kappa=\frac{2}{5}$ and $\alpha= 0.4$ (---), $0.6$ (--$\cdot$--), $0.8$ (-- --), and $1$ ($\cdots$). The circles denote the location of the threshold value ($\beta_0$) below which $\mu_{\min}<0$. \textbf{b} Plot of $\beta_0$ as a function of $\alpha$ for $\kappa=\frac{2}{5}$. The fourth-degree moments diverge in the region below the curve.}\label{fig2}
\end{figure}

The time evolution of the cumulants is characterized by the  eigenvalues  of the matrix $\mathsf{M}$, the asymptotic behavior being governed by the smallest eigenvalue, $\mu_{\min}$. If $\mu_{\min}>0$, the cumulants relax to their finite stationary values $\left[a_{20}^{(0)},a_{02}^{(0)},a_{11}^{(0)},a_{00}^{(1)}\right]^\dagger=-\mathsf{M}^{-1}\cdot\mathsf{L}$. On the other hand, if $\mu_{\min}<0$ then the cumulants diverge.
Figure \ref{fig2}a shows the $\b$-dependence of $\mu_{\min}$ for the same representative values of $\a$ as in Fig.~\ref{fig1}. We observe that, at a given $\a<1$, there exists a threshold value $\b_0$ such that $\mu_{\min}<0$ if $\b<\b_0$. The $\a$-dependence of $\b_0$ is shown in Fig.~\ref{fig2}b. For the pairs $(\a,\b)$ below the curve in Fig.~\ref{fig2}b, the cumulants, and hence the fourth-degree moments, diverge. This divergence is likely   associated with an algebraic high-velocity tail of the VDF, as already present in the case of the IMM
\cite{BMP02,EB02a,EB02b,BK02,KB02,EB02a,EB02b,EB02c,GS07,GS11,SG12}.

If the particles are perfectly smooth ($\b=-1$), the temperature ratio $\theta$ is irrelevant and all the matrix elements vanish except $M_{11}=\frac{1}{120}(1+\a)^2(5+6\a-3\a^2)$, $M_{33}=\frac{1}{20}(1+\a)^2$,  $M_{43}=\frac{1}{36}(1+\a)^2$, $M_{44}=\frac{1}{9}(1+\a)^2$, and $L_1=-\frac{1}{20}(1-\a^2)^2$. This implies that $a_{02}^{(0)}$ is arbitrary, $a_{11}^{(0)}=a_{00}^{(1)}=0$, and $a_{20}^{(0)}=6(1-\a)^2/(5+6\a-3\a^2)$. The latter result agrees with a previous derivation for the IMM \cite{S03}.

On the other hand, in the special case of the Pidduck gas ($\a=\b=1$), the HCS reduces to equilibrium and Eqs.\ \eqref{22ab} yield $\theta=1$, as expected. Moreover, from Eqs.\ \eqref{test_2P} and \eqref{test_3}, one gets $\mathsf{L}=0$, implying that the cumulants \eqref{cumu} vanish in that case.

\begin{figure}[h]
\centering
\includegraphics[width=0.5\textwidth]{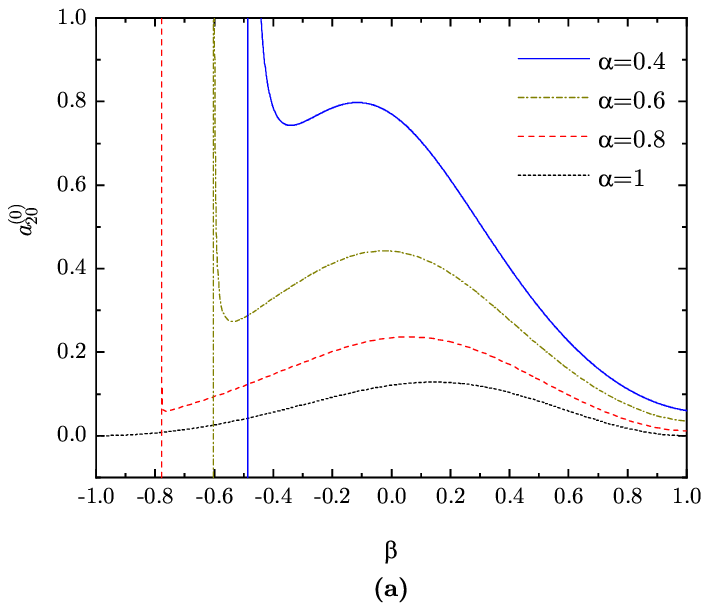}\includegraphics[width=0.5\textwidth]{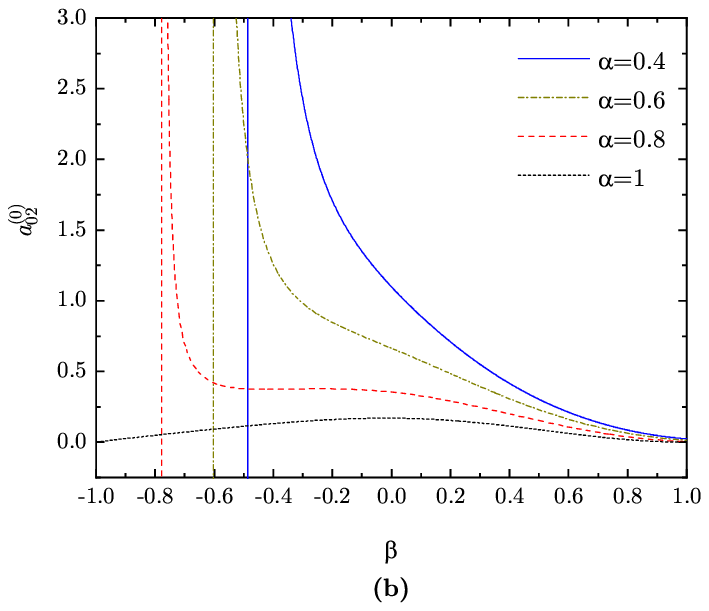}\\
\includegraphics[width=0.5\textwidth]{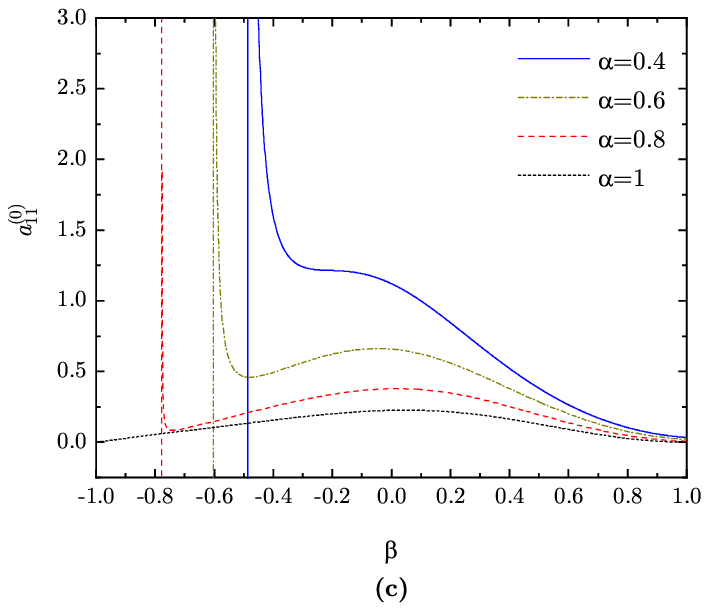}\includegraphics[width=0.5\textwidth]{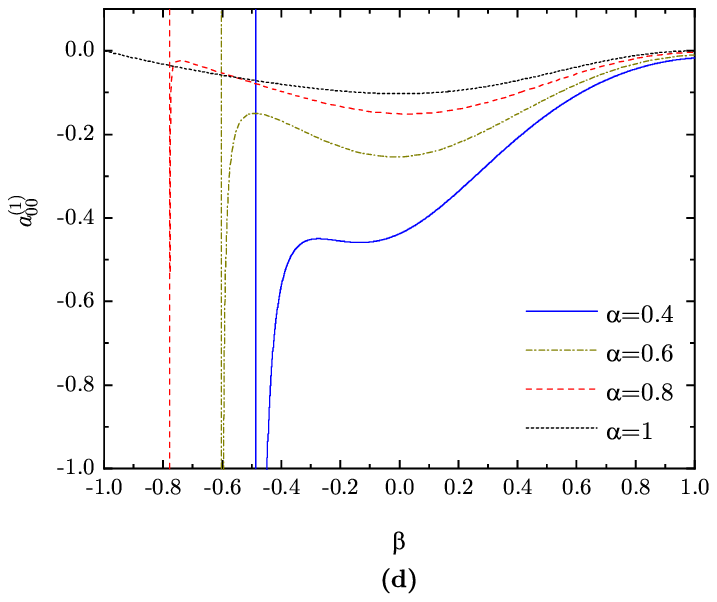}
\caption{Plot of the cumulants \textbf{a} $a_{20}^{(0)}$, \textbf{b} $a_{02}^{(0)}$, \textbf{c} $a_{11}^{(0)}$, and \textbf{d} $a_{00}^{(1)}$ as functions of $\b$ for $\kappa=\frac{2}{5}$ and $\alpha= 0.4$ (---), $0.6$ (--$\cdot$--), $0.8$ (-- --), and $1$ ($\cdots$). The vertical lines denote the asymptotes at $\b=\b_0$.}\label{fig3}
\end{figure}

Apart from the IMM and the equilibrium Pidduck gas,  in the general case with $\a<1$ and $\mid\b\mid<1$, the stationary values of the cumulants are given, as said above, by  $\left[a_{20}^{(0)},a_{02}^{(0)},a_{11}^{(0)},a_{00}^{(1)}\right]^\dagger=-\mathsf{M}^{-1}\cdot\mathsf{L}$, provided that $\b>\b_0$. The $\b$-dependence of the four cumulants are displayed in Fig.~\ref{fig3} for the same choices of $\a$ as in Figs.~\ref{fig1} and \ref{fig2}a.
The positivity of $a_{11}^{(0)}$ implies that high translational velocities are strongly correlated to high angular velocities \cite{SKS11,VSK14,VS15}. In turn, the negative values of $a_{00}^{(1)}$ mean
that quasinormal orientations between the translational and angular velocity vectors tend to be favored against quasiparallel orientations (``lifted-tennis-ball'' effect) \cite{BPKZ07,KBPZ09,RA14,SKS11,VSK14,VS15}.

\section{Conclusions}
\label{sec5}

In this work, we have worked out a simple Maxwell model (IRMM) by keeping the collision rules of inelastic rough hard spheres but, on the other hand, assuming an effective mean-field collision rate
independent of the relative velocity of the colliding pair. The latter assumption allows one to express a collisional moment of degree $k$ as a bilinear combination of velocity moments of degrees
$i\leq k$ and $j\leq k$ with $i+j=k$, as happens in the well-known IMM (smooth particles)
\cite{GS07,SG12}. Nevertheless, the derivation of exact results in the IRMM is much more complicated than in the IMM. Not only are the translational and rotational degrees of freedom entangled, thus increasing the number and structure of the moments, but also the coefficients in the bilinear combinations are functions of the coefficient of normal restitution (as in the IMM) and, additionally, of the coefficient of tangential restitution and the reduced moment of inertia.

Specifically, we have considered the moments of first and second degree, the moments of third degree related to the heat flux, and the isotropic fourth-degree moments.
The structure of the associated collisional moments is displayed in Table \ref{table1}. The number of coefficients is $60$, since in some cases a common coefficient  factorizes
more than one product of moments. The exact expressions of those $60$ coefficients in terms of the two coefficients of restitution ($\a$ and $\b$) and of the reduced moment of inertia ($\ka$)
are presented in Appendix \ref{appB}. We have checked that the coefficients satisfy some consistency tests. First, they reduce to the expressions reported in the literature in the smooth limit ($\b=-1$)
\cite{GS07,SG12}. Next, when particularized to elastic and perfectly rough particles (Pidduck model, $\a=\b=1$), the coefficients obey a number of relations needed to allow the equilibrium VDF to be an exact solution of the model.

The knowledge of the collisional moments stemming from the IRMM is not constrained to a specific physical situation and, thus, it can be exploited in several applications. Here we have applied the results to the HCS. Specifically, the rotational-to-translational temperature ratio has been found [see Eqs.~\eqref{22ab}] and the linear evolution equations for the fourth-degree cumulants have been obtained [see Eq.~\eqref{25}]. Interestingly, we have  found that those cumulants diverge in time if, at a given $\a$, the coefficient of tangential restitution lies below a threshold value $\b_0(\a)$ [see Fig.~\ref{fig2}]. This reflects the existence of an algebraic high-velocity tail of the VDF, reminiscent of the case of the IMM
\cite{BMP02,EB02a,EB02b,BK02,KB02,EB02c}.
If $\b>\b_0(\a)$, the exact stationary  cumulants have been obtained (see Fig.~\ref{fig3}). In general, the departure of the HCS VDF from the equilibrium one is rather strong: (a) the (marginal) translational and angular VDFs exhibit large excess kurtoses, (b) high translational velocities are correlated to high angular velocities, and (c) the translational and angular velocities tend to adopt quasi-normal orientations (``lifted-tennis-ball'' effect).

It must be stressed that, in analogy to the relationship between the IHSM and the IMM, the IRMM is proposed here as a mathematical model and not as an approximation of the IRHSM. In this respect, the predictions obtained analytically from the IRMM do not need to be interpreted strictly as a \emph{quantitative} replacement for the results obtained either numerically or by simulations from the IRHSM. On the other hand, the IRMM provides a tractable toy model that can be used to cleanly unveil nontrivial physical properties, guide the construction of approximations to the IRHSM, and serve as a benchmark for numerical or simulation approaches. Notwithstanding this, a closer contact with the IRHSM can be provided by the augmented IRMM defined by Eq.~\eqref{17}.

Relying on the set of collisional moments displayed in Table \ref{table1} with coefficients given in Appendix \ref{appB}, further applications of the model are envisioned. In particular, the Navier--Stokes transport coefficients can be exactly derived. Preliminary results show the existence of a new ``spin'' viscosity coefficient, which was  overlooked in previous Sonine approximations of the IRHSM \cite{KSG14,MS21a}. In addition, the exact non-Newtonian rheological properties of a granular gas under simple shear flow can also be obtained. The results of both applications will be published elsewhere.

\backmatter

\bmhead{Acknowledgments}

A.S.\ acknowledges financial support from Grant PID2020-112936GB-I00 funded by MCIN/AEI/10.13039/501100011033, and from grants IB20079 and GR21014 funded by Junta de Extremadura (Spain) and by ERDF ``A way of making Europe.''
G.M.K.\ is grateful to the Conselho Nacional de Desenvolvimento Cient\'{i}fico e Tecnol\'{o}gico (CNPq) for financial support through Grant No.\  304054/2019-4.

\bmhead{Data availability}
The datasets employed to generate Figs.~\ref{fig1}--\ref{fig3}  are available from the corresponding author on reasonable request.

\begin{appendices}

\section{Explicit expressions for the coefficients appearing in Table \ref{table1}}
\label{appB}
\setcounter{equation}{0}
The $60$ coefficients appearing in Table \ref{table1} are given by
\bal
\label{B1}
\varphi_{01\mid 01}=&\frac{4\bt}{3\ka},\\
\label{B2}
\chi_{20\mid 20}=&\frac{2}{3}\left[\at\left(1-\at\right)+2\bt\left(1-\bt\right)\right],\quad \chi_{20\mid 02}=-\frac{\bt^2}{3},\\
\label{B3}
\psi_{20\mid 20}=&\frac{2}{15}\left(5\at-2\at^2-6\at\bt+10\bt-7\bt^2\right),\quad  \psi_{20\mid 02}=\frac{\bt^2}{6},\\
\label{B4}
\chi_{02\mid 02}=&\frac{4\bt}{3\ka}\left(1-\frac{\bt}{\ka}\right),\quad \chi_{02\mid 20}=-\frac{16\bt^2}{3\ka^2},\\
\psi_{02\mid 02}=&\frac{2\bt}{15\ka}\left(10-\frac{7\bt}{\ka}\right)
, \quad \psi_{02\mid 20}=\frac{8\bt^2}{3\ka^2},\\
\psi_{11\mid 11}=&\frac{1}{3}\left(\at+2\bt\frac{1+\ka}{\ka}\right),\\
\varphi_{30\mid 30}=&\frac{1}{15}\left(15\at-11\at^2-8\at\bt+30\bt-26\bt^2\right),
\quad
\varphi_{30\mid 12}=-\frac{\bt^2}{6},
\eal
\begin{subequations}
\bal
\varphi_{12\mid 30}=&-\frac{8\bt^2}{3\ka^2},\quad \varphi_{12\mid 12}^{(1)}=\frac{1}{15}\left[5\at+10\bt+\frac{2\bt}{\ka}\left(10-4\at-11\bt-\frac{5\bt}{\ka}\right)\right]
,\\
\varphi_{12\mid 12}^{(2)}=&\frac{2\bt}{15\ka}\left(2\at+3\bt\right)
,\quad \varphi_{12\mid 12}^{(3)}=\frac{4\bt}{3\ka}\left(1-\frac{\bt}{\ka}\right)
,\quad
\varphi_{12\mid 12}^{(4)}=\frac{2\bt^2}{3\ka},
\eal
\end{subequations}
\begin{subequations}
\bal
\overline{\varphi}_{12\mid 12}^{(1)}=&
\frac{1}{15}\left[5\at+10\bt+\frac{\bt}{\ka}\left(20-3\at-7\bt\frac{1+\ka}{\ka}\right)\right]
,\\
 \overline{\varphi}_{12\mid 12}^{(2)}=&\frac{\bt}{15\ka}\left(\at-\bt\frac{1+\ka}{\ka}\right),\quad
 \overline{\varphi}_{12\mid 12}^{(3)}=\frac{\bt}{15\ka}\left(10+5\at+15\bt-\frac{7\bt}{\ka}\right)
 ,\\
 \overline{\varphi}_{12\mid 12}^{(4)}=&-\frac{2\bt^2}{15\ka^2},\quad
 \overline{\varphi}_{12\mid 12}^{(5)}=\frac{\bt}{15\ka}\left(10-5\at-15\bt-\frac{7\bt}{\ka}\right)
 ,
 \eal
 \end{subequations}
\begin{subequations}
    \bal
  \chi_{40\mid 40}^{(1)}=&\frac{2}{15}\Big[10\at-11\at^2+6\at^3-3\at^4+20\bt-26\bt^2+16\bt^3-8\bt^4\nn
  &-4\at\b(2-\a-\b+\a\b)\Big],\\
  \chi_{40\mid 40}^{(2)}=&-\frac{2}{15}\Big[7\at^2-6\at^3+3\at^4+12\bt^2-16\bt^3+8\bt^4  \nn
  &-4\at\b(1+\a+\b-\a\b)\Big],\\
  \chi_{40\mid 40}^{(3)}=&-\frac{4}{15}\Big[2\at^2-6\at^3+3\at^4+7\bt^2-16\bt^3+8\bt^4
    \nn
  &+2\a\b(3-2\a-2\b+2\a\b)],\\
  \chi_{40\mid 04}=&-\frac{\bt^4}{15},\quad
  \chi_{40\mid 22}^{(1)}=-\frac{\bt^2}{15}\left[5-2\at\left(1-\at\right)-8\bt\left(1-\bt\right)\right],\\
  \chi_{40\mid 22}^{(2)}=&\frac{2\bt^2}{15}\left[\at\left(1-\at\right)+4\bt\left(1-\bt\right)\right],
  \eal
\end{subequations}
\begin{subequations}
\bal
 \chi_{04\mid 40}=&-\frac{256\bt^4}{15\ka^4},\quad  \chi_{04\mid 22}^{(1)}=-\frac{16\bt^2}{15\ka^2}\left(5-\frac{8\bt}{\ka}+\frac{8\bt^2}{\ka^2}\right),\\
 \chi_{04\mid 22}^{(2)}=&-\frac{128\bt^3}{15\ka^3}\left(1-\frac{\bt}{\ka}\right),\quad
 \chi_{04\mid 04}^{(1)}=\frac{4\bt}{15\ka}\left(10-\frac{13\bt}{\ka}+\frac{8\bt^2}{\ka^2}-\frac{4\bt^3}{\ka^3}\right),\\
 \chi_{04\mid 04}^{(2)}=&-\frac{8\bt^2}{15\ka^2}\left(3-\frac{4\bt}{\ka}+\frac{2\bt^2}{\ka^2}\right),
 \quad
  \chi_{04\mid 04}^{(3)}=-\frac{4\bt^2}{15\ka^2}\left(7-\frac{16\bt}{\ka}+\frac{8\bt^2}{\ka^2}\right),\\
  \chi_{04\mid 04}^{(4)}=&\frac{8\bt}{15\ka}\left(1-\frac{\b}{\ka}\right)\left(5-\frac{8\bt}{\ka}+\frac{8\bt^2}{\ka^2}\right),
  \eal
\end{subequations}
\begin{subequations}
  \bal
 \chi_{22\mid 40}^{(1)} =&-\frac{8\bt^2}{15\ka^2}\left[5-2\at\left(1-\at\right)-8\bt\left(1-\bt\right)\right],\\
  \chi_{22\mid 40}^{(2)}=&\frac{32\bt^2}{15\ka^2}\left[\at\left(1-\at\right)+4\bt\left(1-\bt\right)\right],\\
  \chi_{22\mid 22}^{(1)}=&\frac{1}{15}\left[5\at(2-\at)+20\bt\frac{1+\ka}{\ka}-2\bt^2\frac{5+22\ka+5\ka^2}{\ka^2}-\frac{8\at(2-\at)\bt}{\ka}
  \right.\nn
  &\left.+\frac{8\at(1-\at)\bt^2}{\ka^2}+32\bt^3\frac{1+\ka}{\ka^2}-\frac{64\bt^4}{\ka^2}\right],\\
 \chi_{22\mid 22}^{(2)}=&-\frac{1}{15}\left[5\at^2+10\bt^2\frac{1+\ka^2}{\ka^2}-\frac{8\at(1-\at)\bt^2}{\ka^2}-\frac{8\at^2\bt}{\ka}
 -32\bt^3\frac{1+\ka}{\ka^2}
  \right.\nn
 &\left.+\frac{64\bt^4}{\ka^2}\right],\\
 \chi_{22\mid 22}^{(3)}=&\frac{4\bt}{15\ka}\left[\at(2-\at)+3\bt-4\bt^2\frac{1+\ka}{\ka}-\frac{\at(1-\at)\bt}{\ka}+\frac{8\bt^3}{\ka}\right],\\
\chi_{22\mid 22}^{(4)}=&-\frac{4\bt}{15\ka}\left[\at^2+4\bt^2\frac{1+\ka}{\ka}+\frac{\at(1-\at)\bt}{\ka}-\frac{8\bt^3}{\ka}\right],\\
 \chi_{22\mid 22}^{(5)}=&\frac{4\bt}{15\ka}\left[2\at(1-\at)+3\bt-8\bt^2\frac{1+\ka}{\ka}-\frac{2\at(1-\at)\bt}{\ka}+\frac{16\bt^3}{\ka}\right],\\
 \chi_{22\mid 22}^{(6)}=&\frac{4\bt}{15\ka}\left[5-4\at(1-\at)-\bt\frac{5+11\ka}{\ka}+16\bt^2\frac{1+\ka}{\ka}
 +\frac{4\at(1-\at)\bt}{\ka}
 -\frac{32\bt^3}{\ka}\right],\\
 \chi_{22\mid 22}^{(7)}=&-\frac{4\bt}{15\ka}\left[\at(1-\at)+4\bt-4\bt^2\frac{1+\ka}{\ka}-\frac{\at(1-\at)\bt}{\ka}+\frac{8\bt^3}{\ka}\right],\\
 \chi_{22\mid 22}^{(8)}=&-\frac{4\bt}{15\ka}\left[\at(1-\at)-\bt-4\bt^2\frac{1+\ka}{\ka}-\frac{\at(1-\at)\bt}{\ka}+\frac{8\bt^3}{\ka}\right],\\
 \chi_{22\mid 22}^{(9)}=&\frac{4\bt}{15\ka}\left[4\at(1-\at)+11\bt-16\bt^2\frac{1+\ka}{\ka}-\frac{4\at(1-\at)\bt}{\ka}+\frac{32\bt^3}{\ka}\right],\\
 \chi_{22\mid 04}^{(1)}=&-\frac{\bt^2}{30}\left(5-\frac{8\bt}{\ka}+\frac{8\bt^2}{\ka^2}\right),\quad
 \chi_{22\mid 04}^{(2)}=\frac{8\bt^3}{15\ka}\left(1-\frac{\bt}{\ka}\right),\\
 \chi_{22\mid 04}^{(3)}=&-\frac{\bt^2}{15}\left(5-\frac{16\bt}{\ka}+\frac{16\bt^2}{\ka^2}\right),
 \eal
\end{subequations}
\begin{subequations}
  \bal
 \overline{\chi}_{22\mid 40}=&-\frac{4\bt^2}{3\ka^2},\\
  \overline{\chi}_{22\mid 22}^{(1)}=&\frac{1}{15}\left[2\at(5-\at)+2\bt(10-3\at)\frac{1+\ka}{\ka}-7\bt^2\frac{(1+\ka)^2}{\ka^2}\right],\\
   \overline{\chi}_{22\mid 22}^{(2)}=&-\frac{1}{15}\left(\at-\bt\frac{1+\ka}{\ka}\right)^2,\quad \overline{\chi}_{22\mid 22}^{(3)}=-\frac{1}{15}\left[\left(\at-\bt\right)^2+\frac{\bt^2}{\ka^2}\right],\\
 \overline{\chi}_{22\mid 22}^{(4)}=&-\frac{1}{15}\left(2\at^2+6\at\bt+7\bt^2\frac{1+\ka^2}{\ka^2}\right),\\
 \overline{\chi}_{22\mid 22}^{(5)}=&\frac{2\bt}{15\ka}\left(10-3\at-7\bt\frac{1+\ka}{\ka}\right),
 \quad  \overline{\chi}_{22\mid 22}^{(6)}=\frac{2\bt}{15\ka}\left(\at-\bt\frac{1+\ka}{\ka}\right),\\
 \overline{\chi}_{22\mid 22}^{(7)}=&-\frac{2\bt}{15\ka}\left(\at+4\bt\right),
 \quad \overline{\chi}_{22\mid 22}^{(8)}=\frac{2\bt}{15\ka}\left(4\at+11\bt\right),\\
 \overline{\chi}_{22\mid 22}^{(9)}=&-\frac{2\bt}{15\ka}\left(\at-\bt\right),
 \quad \overline{\chi}_{22\mid 04}=-\frac{\bt^2}{12}.
 \eal
\end{subequations}

\section{Consistency tests}
\label{appC}

{\renewcommand{\arraystretch}{1.8}
\begin{table}
\begin{center}
\caption{Coefficients associated with the collisional moments of first, second, and third degree in Table \ref{table1} in the special cases of (i) inelastic and perfectly smooth particles ($\a<1$, $\b=-1$) and (ii) elastic and perfectly rough particles ($\a=\b=1$).}\label{table2}
\begin{tabular}{@{}ccc@{}}
\toprule
Coefficient&(i) Inelastic and perfectly smooth  &(ii) Elastic and perfectly rough\\
&($\a<1$, $\b=-1$)&($\a=\b=1$)\\
\midrule
$\varphi_{01\mid 01}$&$0$&$\displaystyle{\frac{4}{3(1+\ka)}}$\\
$\chi_{20\mid 20}$&$\displaystyle{\frac{1-\alpha^2}{6}}$&$\displaystyle{\frac{4\ka}{3(1+\ka)^2}}$\\
$\chi_{20\mid 02}$&$0$&$\displaystyle{-\frac{\ka}{4}\chi_{20\mid 20}}$\\
$\psi_{20\mid 20}$&$\displaystyle{\frac{(1+\a)(4-\a)}{15}}$&$\displaystyle{\frac{2(3+10\ka)}{15(1+\ka)^2}}$\\
$\psi_{20\mid 02}$&$0$&$\displaystyle{\frac{\ka}{8}\chi_{20\mid 20}}$\\
$\chi_{02\mid 02}$&$0$&$\chi_{20\mid 20}$\\
$\chi_{02\mid 20}$&$0$&$\displaystyle{-\frac{4}{\ka}\chi_{20\mid 20}}$\\
$
\psi_{02\mid 02}$&$0$&$\psi_{20\mid 20}$\\
$
\psi_{02\mid 20}$&$0$&$\displaystyle{\frac{2}{\ka}\chi_{20\mid 20}}$\\
$
\psi_{11\mid 11}$&$\displaystyle{\frac{1+\a}{6}}$&$1$\\
$
\varphi_{30\mid 30}$&$\displaystyle{\frac{(1+\a)(19-11\a)}{60}}$&$\displaystyle{\frac{2(2+15\ka)}{15(1+\ka)^2}}$\\
$
\varphi_{30\mid 12}$&$0$&$\displaystyle{-\frac{\ka}{8}\chi_{20\mid 20}}$\\
$
\varphi_{12\mid 30}$&$0$&$\displaystyle{-\frac{2}{\ka}\chi_{20\mid 20}}$\\
$
\varphi_{12\mid 12}^{(1)}$&$\psi_{11\mid 11}$&$\displaystyle{\frac{7+10\ka+15\ka^2}{15(1+\ka)^2}}$\\
$
\varphi_{12\mid 12}^{(2)}$&$0$&$\displaystyle{\frac{2(2+5\ka)}{15(1+\ka)^2}}$\\
$
\varphi_{12\mid 12}^{(3)}$&$0$&$\displaystyle{\chi_{20\mid 20}}$\\
$
\varphi_{12\mid 12}^{(4)}$&$0$&$\displaystyle{\frac{1}{2}\chi_{20\mid 20}}$\\
$
\overline{\varphi}_{12\mid 12}^{(1)}$&$\psi_{11\mid 11}$&$1$\\
$
\overline{\varphi}_{12\mid 12}^{(2)}$&$0$&$0$\\
$
\overline{\varphi}_{12\mid 12}^{(3)}$&$0$&$\displaystyle{\frac{2(4+15\ka)}{15(1+\ka)^2}}$\\
$
\overline{\varphi}_{12\mid 12}^{(4)}$&$0$&$\displaystyle{-\frac{1}{10\ka}\chi_{20\mid 20}}$\\
$
\overline{\varphi}_{12\mid 12}^{(5)}$&$0$&$\displaystyle{-\frac{2(1+5\ka)}{15(1+\ka)^2}}$\\
\botrule
\end{tabular}
\end{center}
\end{table}
}

{\renewcommand{\arraystretch}{1.8}
\begin{table}
\begin{center}
\caption{Coefficients associated with the collisional moments $\mathcal{J}_{\text{M}}[V^4]$ and $\mathcal{J}_{\text{M}}[\omega^4]$ in Table \ref{table1} in the special cases of (i) inelastic and perfectly smooth particles ($\a<1$, $\b=-1$) and (ii) elastic and perfectly rough particles ($\a=\b=1$).}\label{table3}
\begin{tabular}{@{}ccc@{}}
\toprule
Coefficient&(i) Inelastic and perfectly smooth  &(ii) Elastic and perfectly rough\\
&($\a<1$, $\b=-1$)&($\a=\b=1$)\\
\midrule
$\chi_{40\mid 40}^{(1)}$&$\displaystyle{\frac{(1+\a)(45-29\a+3\a^2-3\a^3)}{120}}$&$\displaystyle{\frac{4(1+12\ka+17\ka^2+10\ka^3)}{15(1+\ka)^4}}$\\
$
\chi_{40\mid 40}^{(2)}$&$\displaystyle{-\frac{(1+\a)^2(19-6\a+3\a^2)}{120}}$&$\displaystyle{-\frac{8(1+2\ka+3\ka^2)}{15(1+\ka)^4}}$\\
$
\chi_{40\mid 40}^{(3)}$&$\displaystyle{\frac{(1+\a)^2(1+6\a-3\a^2)}{60}}$&$\displaystyle{\frac{4(1+2\ka-7\ka^2)}{15(1+\ka)^4}}$\\
$
\chi_{40\mid 04}$&$0$&$\displaystyle{-\frac{\ka^4}{15(1+\ka)^4}}$\\
$
\chi_{40\mid 22}^{(1)}$&$0$&$\displaystyle{-\frac{\ka^2(5+2\ka+5\ka^2)}{15(1+\ka)^4}}$\\
$
\chi_{40\mid 22}^{(2)}$&$0$&$\displaystyle{-\frac{8}{\ka}\chi_{40\mid 04}}$\\
$
\chi_{04\mid 40}$&$0$&$\displaystyle{\frac{256}{\ka^4}\chi_{40\mid 04}}$\\
$
\chi_{04\mid 22}^{(1)}$&$0$&$\displaystyle{\frac{16}{\ka^2}\chi_{40\mid 22}^{(1)}}$\\
$
\chi_{04\mid 22}^{(2)}$&$0$&$\displaystyle{\frac{128}{\ka^3}\chi_{40\mid 04}}$\\
$
\chi_{04\mid 04}^{(1)}$&$0$&$\chi_{40\mid 40}^{(1)}$\\
$
\chi_{04\mid 04}^{(2)}$&$0$&$\chi_{40\mid 40}^{(2)}$\\
$
\chi_{04\mid 04}^{(3)}$&$0$&$\chi_{40\mid 40}^{(3)}$\\
$
\chi_{04\mid 04}^{(4)}$&$0$&$\displaystyle{-\frac{8}{\ka}\chi_{40\mid 22}^{(1)}}$\\
\botrule
\end{tabular}
\end{center}
\end{table}
}

{\renewcommand{\arraystretch}{1.8}
\begin{table}
\begin{center}
\caption{Coefficients associated with the collisional moments $\mathcal{J}_{\text{M}}[V^2\omega^2]$ and $\mathcal{J}_{\text{M}}[(\bV\cdot\bw)^2]$ in Table \ref{table1} in the special cases of (i) inelastic and perfectly smooth particles ($\a<1$, $\b=-1$) and (ii) elastic and perfectly rough particles ($\a=\b=1$).}\label{table4}
\begin{tabular}{@{}ccc@{}}
\toprule
Coefficient&(i) Inelastic and perfectly smooth  &(ii) Elastic and perfectly rough\\
&($\a<1$, $\b=-1$)&($\a=\b=1$)\\
\midrule
$
\chi_{22\mid 40}^{(1)}$&$0$&$\displaystyle{\frac{8}{\ka^2}\chi_{40\mid 22}^{(1)}}$\\
$
\chi_{22\mid 40}^{(2)}$&$0$&$\displaystyle{-\frac{128}{\ka^3}\chi_{40\mid 04}}$\\
$
\chi_{22\mid 22}^{(1)}$&$\displaystyle{\frac{(1+\a)(3-\a)}{12}}$&$\displaystyle{\frac{7+44\ka+18\ka^2+60\ka^3+15\ka^4}{15(1+\ka)^4}}$\\
$
\chi_{22\mid 22}^{(2)}$&$\displaystyle{-\frac{(1+\a)^2}{12}}$&$\displaystyle{-\frac{7-16\ka+26\ka^2+15\ka^4}{15(1+\ka)^4}}$\\
$
\chi_{22\mid 22}^{(3)}$&$0$&$\displaystyle{\frac{4(1+2\ka+9\ka^2)}{15(1+\ka)^4}}$\\
$
\chi_{22\mid 22}^{(4)}$&$0$&$\displaystyle{-\frac{4(1+7\ka+3\ka^2+5\ka^3)}{15(1+\ka)^4}}$\\
$
\chi_{22\mid 22}^{(5)}$&$0$&$\displaystyle{-\frac{4\ka(5-6\ka+5\ka^2)}{15(1+\ka)^4}}$\\
$
\chi_{22\mid 22}^{(6)}$&$0$&$-2\chi_{22\mid 22}^{(5)}$\\
$
\chi_{22\mid 22}^{(7)}$&$0$&$\displaystyle{\frac{32}{\ka^2}\chi_{40\mid 04}}$\\
$
\chi_{22\mid 22}^{(8)}$&$0$&$\displaystyle{-\frac{4}{\ka}\chi_{40\mid 22}^{(1)}}$\\
$
\chi_{22\mid 22}^{(9)}$&$0$&$\displaystyle{-\frac{4\ka(5-22\ka+5\ka^2)}{15(1+\ka)^4}}$\\
$
\chi_{22\mid 04}^{(1)}$&$0$&$\displaystyle{\frac{1}{2}\chi_{40\mid 22}^{(1)}}$\\
$
\chi_{22\mid 04}^{(2)}$&$0$&$\displaystyle{-\frac{8}{\ka}\chi_{40\mid 04}}$\\
$
\chi_{22\mid 04}^{(3)}$&$0$&$\displaystyle{\frac{\ka}{4}\chi_{22\mid 22}^{(5)}}$\\
$
\overline{\chi}_{22\mid 40}$&$0$&$\displaystyle{-\frac{1}{\ka}}\chi_{20\mid 20}$\\
$
\overline{\chi}_{22\mid 22}^{(1)}$&$\displaystyle{\frac{(1+\a)(9-\a)}{30}}$&$1$\\
$
\overline{\chi}_{22\mid 22}^{(2)}$&$\displaystyle{\frac{1}{5}\chi_{22\mid 22}^{(2)}}$&$0$\\
$
\overline{\chi}_{22\mid 22}^{(3)}$&$\displaystyle{\frac{1}{5}\chi_{22\mid 22}^{(2)}}$&$\displaystyle{-\frac{1}{10\ka}\chi_{20\mid 20}}$\\
$
\overline{\chi}_{22\mid 22}^{(4)}$&$\displaystyle{\frac{2}{5}\chi_{22\mid 22}^{(2)}}$&$\displaystyle{-\frac{9+10\ka+15\ka^2}{15(1+\ka)^2}}$\\
$
\overline{\chi}_{22\mid 22}^{(5)}$&$0$&$0$\\
$
\overline{\chi}_{22\mid 22}^{(6)}$&$0$&$0$\\
$
\overline{\chi}_{22\mid 22}^{(7)}$&$0$&$\overline{\varphi}_{12\mid 12}^{(5)}$\\
$
\overline{\chi}_{22\mid 22}^{(8)}$&$0$&$\overline{\varphi}_{12\mid 12}^{(3)}$\\
$
\overline{\chi}_{22\mid 22}^{(9)}$&$0$&$\displaystyle{-\frac{1}{10\ka}\chi_{20\mid 20}}$\\
$
\overline{\chi}_{22\mid 04}$&$0$&$\displaystyle{-\frac{\ka}{16}\chi_{20\mid 20}}$\\
\botrule
\end{tabular}
\end{center}
\end{table}
}

Tables \ref{table2}--\ref{table4} display the expressions of the $60$ coefficients in two interesting situations: (i) inelastic and perfectly smooth particles (i.e., the IMM, $\a<1$, $\b=-1$) and (ii) elastic and perfectly rough particles (i.e., the Pidduck model, $\a=\b=1$).
Let us use both situations (i) and (ii) as tests of the results.

\subsection{Inelastic Maxwell model (IMM)}
In case (i), $45$ out of the $60$ coefficients vanish. First, the angular velocities are unaffected by collisions and thus
$\mathcal{J}_{\text{M}}[\Psi_{0k_2}]=0$ because $\Psi_{0k_2}(\bxi)$ is a function of $\bw$ only. This implies that the $12$ coefficients of the form $Y_{0k_2\mid \ell_1\ell_2}$ identically vanish.

Also, since $\Psi_{k_10}(\bxi)$ is a function of $\bV$ only, $\mathcal{J}_{\text{M}}[\Psi_{k_10}]$ cannot be coupled to moments involving the angular velocity, so that $Y_{k_10\mid \ell_1\ell_2}=0$ if $\ell_2\neq 0$ ($6$ coefficients).

Next, the angular and translational velocities are uncorrelated and, therefore, $\mathcal{J}_{\text{M}}[\Psi_{k_1k_2}]=\langle\Psi_{k_2}^{(2)}(\bw)\rangle \mathcal{J}_{\text{M}}[\Psi_{k_1}^{(1)}]$ and $\langle\Psi_{k_1k_2}(\bxi)\rangle=\langle\Psi_{k_1}^{(1)}(\bV)\rangle\langle\Psi_{k_2}^{(2)}(\bw)\rangle$. This justifies the vanishing of the remaining $27$ coefficients, as well as the relations
\begin{subequations}
\beq
\label{test_1}
\chi_{22\mid 22}^{(1)}+\chi_{22\mid 22}^{(2)}=\chi_{20\mid 20},
\eeq
\beq
\label{test_2}
\overline{\chi}_{22\mid 22}^{(2)}+\overline{\chi}_{22\mid 22}^{(3)}=\frac{1}{3}\left(\chi_{20\mid 20}-\psi_{20\mid 20}\right),
\quad
\overline{\chi}_{22\mid 22}^{(1)}+\overline{\chi}_{22\mid 22}^{(4)}=\psi_{20\mid 20}.
\eeq
\end{subequations}
Equations \eqref{test_1} and \eqref{test_2} imply $\mathcal{J}_{\text{M}}[V^2\omega^2]=\medio{\omega^2}\mathcal{J}_{\text{M}}[V^2]$ and $\mathcal{J}_{\text{M}}[(\bV\cdot\bw)^2]=\medio{\bw\bw}:\mathcal{J}_{\text{M}}[\bV\bV]$, respectively.

Finally, the coefficients $\chi_{20\mid 20}$, $\psi_{20\mid 20}$, $\varphi_{30\mid 30}$, $\chi_{40\mid 40}^{(1)}$, $\chi_{40\mid 40}^{(2)}$, and $\chi_{40\mid 40}^{(3)}$ agree with previous results for the IMM
\cite{S03,SG07,GS07,GS11,SG12,G19}.

\subsection{Pidduck model}
In case (ii) the total energy is conserved by collisions, i.e., $\mathcal{J}_{\text{M}}[mV^2+I\omega^2]=0$. This is guaranteed by the relations
\beq
\label{test_2P}
\chi_{20\mid 20}+\frac{\ka}{4}\chi_{02\mid 20}=0,\quad \chi_{20\mid 02}+\frac{\ka}{4}\chi_{02\mid 02}=0.
\eeq
Other tests in case (ii) correspond to the fact that $\mathcal{J}_{\text{M}}[\Psi]=0$ for any $\Psi(\bxi)$ if the system is at equilibrium, in which case all the anisotropic moments vanish and $\langle\omega^2\rangle=\frac{m}{I}\medio{V^2}=\frac{4}{\ka\sigma^2}\medio{V^2}$, $\medio{V^2\omega^2}=3\medio{\left(\bV\cdot\bw\right)^2}=\medio{V^2}\medio{\omega^2}$, $\medio{V^4}=\frac{5}{3}\medio{V^2}^2$, and $\medio{\omega^4}=\frac{5}{3}\medio{\omega^2}^2$. Therefore, according to Table \ref{table1}, one should have
\begin{subequations}
\label{test_3}
\beq
\chi_{20\mid 20}+\frac{4}{\ka}\chi_{20\mid 02}=0,\quad \chi_{02\mid 20}+\frac{4}{\ka}\chi_{02\mid 02}=0,
\eeq
\beq
{5}\chi_{40\mid 40}^{(1)}+3\chi_{40\mid 40}^{(2)}+\chi_{40\mid 40}^{(3)}+\frac{40}{\ka}\chi_{40\mid 22}^{(1)}+\frac{160}{\ka^2}\chi_{40\mid 04}=0,
\eeq
\beq
5\chi_{04\mid 40}+\frac{20}{\ka}\chi_{04\mid 22}^{(1)}+\frac{8}{\ka^2}\left[5\chi_{04\mid 04}^{(1)}+3\chi_{04\mid 04}^{(2)}+\chi_{04\mid 04}^{(3)}\right]=0,
\eeq
\bal
8\chi_{22\mid 40}^{(1)}+&\chi_{22\mid 40}^{(2)}+\frac{4}{\ka}\left[3\chi_{22\mid 22}^{(1)}+3\chi_{22\mid 22}^{(2)}+\chi_{22\mid 22}^{(3)}+\chi_{22\mid 22}^{(4)}\right]\nn
+&\frac{16}{\ka^2}\left[8\chi_{22\mid 04}^{(1)}+\chi_{22\mid 04}^{(2)}\right]=0,
\eal
\beq
\overline{\chi}_{22\mid 40}+\frac{2}{\ka}\left[\overline{\chi}_{22\mid 22}^{(1)}+3\overline{\chi}_{22\mid 22}^{(2)}+3\overline{\chi}_{22\mid 22}^{(3)}+\overline{\chi}_{22\mid 22}^{(4)}\right]+\frac{16}{\ka^2}\overline{\chi}_{22\mid 04}=0.
\eeq
From the expressions in Tables \ref{table2}--\ref{table4} one can check that Eqs.\ \eqref{test_3} are indeed satisfied.
\end{subequations}

\end{appendices}






\end{document}